
\def\J{$J/\psi$}
\def\j{J/\psi}
\def\P{$\psi'$}
\def\p{\psi'}
\def\U{$\Upsilon$}
\def\u{\Upsilon}
\def\c{c{\bar c}}
\def\b{b{\bar b}}
\def\F{$\Phi$}
\def\f{\Phi}

\def\t{\tau}

\def\q{q{\bar q}}
\def\Q{Q{\bar Q}}

\def\l{\lambda}
\def\e{\epsilon}

\def\lsim{\raise0.3ex\hbox{$<$\kern-0.75em\raise-1.1ex\hbox{$\sim$}}}
\def\gsim{\raise0.3ex\hbox{$>$\kern-0.75em\raise-1.1ex\hbox{$\sim$}}}

\newcount\REFERENCENUMBER\REFERENCENUMBER=0
\def\REF#1{\expandafter\ifx\csname RF#1\endcsname\relax
               \global\advance\REFERENCENUMBER by 1
               \expandafter\xdef\csname RF#1\endcsname
                   {\the\REFERENCENUMBER}\fi}
\def\reftag#1{\expandafter\ifx\csname RF#1\endcsname\relax
               \global\advance\REFERENCENUMBER by 1
               \expandafter\xdef\csname RF#1\endcsname
                      {\the\REFERENCENUMBER}\fi
             \csname RF#1\endcsname\relax}
\def\ref#1{\expandafter\ifx\csname RF#1\endcsname\relax
               \global\advance\REFERENCENUMBER by 1
               \expandafter\xdef\csname RF#1\endcsname
                      {\the\REFERENCENUMBER}\fi
             [\csname RF#1\endcsname]\relax}
\def\refto#1#2{\expandafter\ifx\csname RF#1\endcsname\relax
               \global\advance\REFERENCENUMBER by 1
               \expandafter\xdef\csname RF#1\endcsname
                      {\the\REFERENCENUMBER}\fi
           \expandafter\ifx\csname RF#2\endcsname\relax
               \global\advance\REFERENCENUMBER by 1
               \expandafter\xdef\csname RF#2\endcsname
                      {\the\REFERENCENUMBER}\fi
             [\csname RF#1\endcsname--\csname RF#2\endcsname]\relax}
\def\refs#1#2{\expandafter\ifx\csname RF#1\endcsname\relax
               \global\advance\REFERENCENUMBER by 1
               \expandafter\xdef\csname RF#1\endcsname
                      {\the\REFERENCENUMBER}\fi
           \expandafter\ifx\csname RF#2\endcsname\relax
               \global\advance\REFERENCENUMBER by 1
               \expandafter\xdef\csname RF#2\endcsname
                      {\the\REFERENCENUMBER}\fi
            [\csname RF#1\endcsname,\csname RF#2\endcsname]\relax}
\def\refss#1#2#3{\expandafter\ifx\csname RF#1\endcsname\relax
               \global\advance\REFERENCENUMBER by 1
               \expandafter\xdef\csname RF#1\endcsname
                      {\the\REFERENCENUMBER}\fi
           \expandafter\ifx\csname RF#2\endcsname\relax
               \global\advance\REFERENCENUMBER by 1
               \expandafter\xdef\csname RF#2\endcsname
                      {\the\REFERENCENUMBER}\fi
           \expandafter\ifx\csname RF#3\endcsname\relax
               \global\advance\REFERENCENUMBER by 1
               \expandafter\xdef\csname RF#3\endcsname
                      {\the\REFERENCENUMBER}\fi
[\csname RF#1\endcsname,\csname RF#2\endcsname,\csname
RF#3\endcsname]\relax}
\def\refand#1#2{\expandafter\ifx\csname RF#1\endcsname\relax
               \global\advance\REFERENCENUMBER by 1
               \expandafter\xdef\csname RF#1\endcsname
                      {\the\REFERENCENUMBER}\fi
           \expandafter\ifx\csname RF#2\endcsname\relax
               \global\advance\REFERENCENUMBER by 1
               \expandafter\xdef\csname RF#2\endcsname
                      {\the\REFERENCENUMBER}\fi
            [\csname RF#1\endcsname,\csname RF#2\endcsname]\relax}
\def\Ref#1{\expandafter\ifx\csname RF#1\endcsname\relax
               \global\advance\REFERENCENUMBER by 1
               \expandafter\xdef\csname RF#1\endcsname
                      {\the\REFERENCENUMBER}\fi
             [\csname RF#1\endcsname]\relax}
\def\Refto#1#2{\expandafter\ifx\csname RF#1\endcsname\relax
               \global\advance\REFERENCENUMBER by 1
               \expandafter\xdef\csname RF#1\endcsname
                      {\the\REFERENCENUMBER}\fi
           \expandafter\ifx\csname RF#2\endcsname\relax
               \global\advance\REFERENCENUMBER by 1
               \expandafter\xdef\csname RF#2\endcsname
                      {\the\REFERENCENUMBER}\fi
            [\csname RF#1\endcsname--\csname RF#2\endcsname]\relax}
\def\Refand#1#2{\expandafter\ifx\csname RF#1\endcsname\relax
               \global\advance\REFERENCENUMBER by 1
               \expandafter\xdef\csname RF#1\endcsname
                      {\the\REFERENCENUMBER}\fi
           \expandafter\ifx\csname RF#2\endcsname\relax
               \global\advance\REFERENCENUMBER by 1
               \expandafter\xdef\csname RF#2\endcsname
                      {\the\REFERENCENUMBER}\fi
        [\csname RF#1\endcsname,\csname RF#2\endcsname]\relax}
\def\refadd#1{\expandafter\ifx\csname RF#1\endcsname\relax
               \global\advance\REFERENCENUMBER by 1
               \expandafter\xdef\csname RF#1\endcsname
                      {\the\REFERENCENUMBER}\fi \relax}

%

\def\NP{{ Nucl.\ Phys.\ }}
\def\PL{{ Phys.\ Lett.\ }}
\def\PR{{ Phys.\ Rev.\ }}

\def\PRL{{ Phys.\ Rev.\ Lett.\ }}

\def\ZP{{ Z.\ Phys.\ }}

%
\magnification=1200
\hsize=15.2truecm
\vsize=21.9truecm
\baselineskip=13pt
\hoffset= 0.6 truecm
\voffset= 1 truecm
\nopagenumbers
\def\la{\Lambda_{\rm QCD}}
\def\l{\lambda}
\def\S{${\tilde \sigma}_{\c}$}
\def\s{\tilde{\sigma}_{\c}}
\def\Q{$Q{\bar Q}$}
\def\q{q{\bar q}}
\def\la{\Lambda_{\rm QCD}}
\def\F{$\Phi$}
{}~
\bigskip
\centerline{\bf COLOUR DECONFINEMENT AND QUARKONIUM DISSOCIATION}
\bigskip\bigskip
\centerline{D. KHARZEEV and H. SATZ}
\bigskip
\centerline{\sl Theory Division, CERN, CH-1211 Geneva, Switzerland}
\par
\centerline{\sl and}
\par
\centerline{\sl Fakult\"at f\"ur Physik, Universit\"at Bielefeld,
D-33501 Bielefeld, Germany}
\vskip 1.5 truecm
\centerline{CONTENTS}
\medskip
\leftskip=1.5truecm
\item{1.}{Introduction \hfill \break
          1.1 Preview \hfill\break
          1.2 Probing Colour Deconfinement}
\item{2.}{Quarkonium Production in Hadron-Hadron Collisions\hfill\break
          2.1 Colour Evaporation \hfill\break
          2.2 Quarkonium Production: Theory and Data}
\item{3.}{Quarkonium Evolution and Hadron-Nucleus
          Collisions \hfill\break
          3.1 Parton Fusion and Colour Evaporation in Nuclear
          Matter\hfill\break
          3.2 Quantum-Mechanical Interference and Nuclear Shadowing
          \hfill\break
          3.3 Momentum Loss of a Colour Charge in Nuclear Matter
          \hfill\break
          3.4 Charmonium Interactions in Nuclear Matter
          \hfill\break
          3.5 EMC Effect Modifications}
\item{4.}{The Theory of Quarkonium-Hadron Interactions\hfill\break
          4.1 The Short-Distance Analysis of Quarkonium-Hadron
          Interactions \hfill\break
          4.2 Non-Perturbative Quarkonium Dissociation \hfill\break
          4.3 The Experimental Study of Charmonium-Hadron Interactions}
\item{5.}{Quarkonium Dissociation in Confined and Deconfined Media
          \hfill\break
          5.1 The Parton Structure of Confined vs. Deconfined Matter
          \hfill\break
          5.2 Quarkonium Dissociation by Deconfined Gluons
          \hfill\break
          5.3 Charmonium Survival in an Expanding Medium}
\item{6.}{Pre-Equilibrium Deconfinement \hfill\break
          6.1 Shadowing and \J~Suppression in Nucleus-Nucleus Collisions
          \hfill\break
          6.2 Deconfined Gluons and QGP}
\item{7.}{Conclusions}
\bigskip\vfill
\leftskip=0truecm
\eject
\noindent{\bf 1.\ Introduction}
\medskip
The crucial tool in the search for the quark-gluon plasma (QGP) is a
probe to test if the strongly interacting medium produced in nuclear
collisions consists of confined or deconfined quarks and gluons.
In this survey, we will show how quarkonia can be used as such a tool.
First of all, this requires an understanding of the dynamics of
quarkonium production
in the absence of a thermal medium, i.e., in hadron-hadron collisions.
The theory of quarkonium production can today be tested on a variety of
different states ($\j,~\chi_c,~\p,~\u,~\u',~\u''$); it is confirmed by
data over a vast range of collision energies, up to 1.8 TeV. Given
quarkonium production dynamics, we then have to address three distinct
problems to obtain a viable probe. What are the effects of
confined and of deconfined matter on the production of the
different quarkonium states? How can pre-equilibrium phenomena
affect the production? We begin with a short summary of the answers
which we shall obtain to these questions.
\medskip\noindent
{\sl 1.1~~Preview}
\medskip
To be specific, we consider charmonium states; but the
arguments are in general applicable to bottonium states as well.
The theory of charmonium-hadron interactions predicts that the tightly
bound charmonium ground state \J~cannot be dissociated by hadronic
matter at temperatures below 0.5 GeV. The resulting transparency of
confined media to \J's can be tested experimentally in nuclear
matter, by studying \J-production with a $Pb$-beam incident on a
hydrogen or deuterium target. In contrast, a quark-gluon plasma contains
deconfined gluons, and these are hard enough to break up a physical \J.
\par
Hadron-nucleus collisions also provide the information needed to
eliminate possible initial state effects. By comparing different hard
processes with and without possible final state interactions, the
role of the nuclear medium on the initial state can be clarified and the
final state effects can be singled out.
\par
Combining hadron-nucleus and nucleus-nucleus studies to determine
the effects of confined and deconfined systems on quarkonium
production, we can thus test colour deconfinement. In a nutshell: once
initial state effects are removed, confined systems have no effect on
the \J, but suppress the \P; once deconfinement sets in, both \J~and
\P are suppressed, but the \P~more strongly.
\medskip\noindent
{\sl 1.2~~Probing Colour Deconfinement}
\medskip
After this short preview of things to come, we return to the general
problem of establishing quark-gluon plasma formation.
In high energy nuclear collisions, two beams of partons collide; the
partons are initially confined to the colliding nucleons. This
confinement can be checked, e.g.,\ by studying primary high mass
Drell-Yan dilepton production and observing, except for possible
nuclear shadowing effects, the same
parton distribution functions as in deep inelastic lepton-nucleon
collisions. After the primary collision, we expect abundant
multiple interactions, leading to a rapid increase of entropy, quick
thermalisation and hence the production of strongly interacting matter.
The fundamental question is whether confinement survives this
thermalisation. If it does, we have hadronic matter -- if not, a
quark-gluon plasma. We expect confinement to be lost if the
parton density sufficiently surpasses that present in a
hadron-hadron interaction, so that partons can no longer be
assigned to specific hadrons. How can we check if this has happened?
\par
The QGP is a {\sl dense} system of {\sl deconfined} quarks and gluons.
Its density is in fact the reason for deconfinement:
in a sufficiently dense medium, the long-range confining forces
become screened, so that only short-range ($\ll \Lambda_{\rm QCD}^{-1}$)
interactions between quarks and gluons remain. To study such a medium
and determine its nature, we therefore need probes which are {\sl hard}
enough to resolve the short sub-hadronic scales and which can distinguish
between {\sl confined} and {\sl deconfined} quarks and gluons.
In addition, the probe must survive the subsequent evolution of the
medium; therefore it certainly cannot be in equilibrium with the later
stages of matter. Two hard, strongly interacting signals produced before
equilibration and distinct from the medium have been proposed as
probes for confinement/deconfinement:
\item{--}{{\sl heavy} quark-antiquark resonances (charmonium, bottonium)
\refs{Matsui}{KS3}, and}
\item{--}{{\sl hard} quarks or gluons (jets) \refs{Bj}{Miklos}.}
\par\noindent
In the following, we will comment only briefly on jets, in
order to point out the relation between the two probes, and
then concentrate on quarkonium production.
Quarkonium and jet production are rather well understood in
hadron-hadron collisions, where they are accounted for in terms of
perturbative QCD and hadronic parton distribution functions
\refs{Jets}{Quarko}.
In both cases, the initially formed state ($Q{\bar Q}$, $q$ or $g$) is
in general coloured, and it has an intrinsic mass or momentum scale
much larger than $\Lambda_{\rm QCD}$. For jets, this is also the state
to be used as probe, since the behaviour of a fast colour charge
passing through confined matter differs from that in a deconfined medium
\ref{Miklos/XNW}. In confined matter, the colour charge loses energy as
it passes from one hadron to the next through the ``interhadronic"
vacuum, and the energy loss is determined by the string tension
$\sigma$ acting on the colour charge as it leaves the field of a hadron
\ref{KS1}. In a hot deconfined medium, the crucial quantities are
the colour screening radius and the mean free path; these determine
with how many other charges the passing colour charge interacts and
how much energy it loses per unit length
\refs{Gyu-Wa}{Baier}. The fate of a colour charge in
the transition region between these two limits is still quite
uncertain.
\par
For fast quarkonia, the situation is similar; they will pass through
the medium while still in a coloured state \ref{KS1}, and hence they
can
be used as probe in the same way as jets. In addition, however, we can
consider slow quarkonia, which have become full physical resonances
within a hadronic volume around the \Q~formation point and thus
traverse the medium as colour singlets. Since the intrinsic spatial
scales of \J~and \U, determined by the heavy quark masses and the
binding energies, nevertheless remain much smaller than the hadronic
size $\Lambda_{\rm QCD}^{-1}$, they interact only with the partons
within a big, light hadron and not with the hadron as a whole. They are
thus able to probe the partonic state of any medium. In particular,
they are essentially unaffected by the soft gluons which make up
confined matter, while the hard gluons present in a QGP will dissociate
them \ref{KS3}.
\par
For both quarkonia and jets, thermal production in the expected
temperature range ($T~\lsim~0.5$ GeV) is excluded by the mass or
momentum scales involved; we can therefore be quite sure that such
signals were produced prior to QGP formation. They will also not
reach an equilibrium with later stages of the medium.
Hard jets and fast quarkonia require too much of an energy loss for
this, while slow quarkonia, as noted, are either dissociated
or not affected by the medium.
\par
For both proposed probes, initial state nuclear effects can occur
before QGP formation. Primary quarks and gluons may undergo multiple
scattering or experience shadowing in the nucleus {\sl before}
they interact to form a \Q~pair or a hard transverse parton. These
effects have to be understood and taken into account before any QGP
analysis \refs{GS1}{KS2}. It is therefore necessary to
study them in processes which are not effected by the subsequent
medium, such as
the production of hard direct photons \ref{Cley} or of high
mass Drell-Yan dileptons \ref{VesaHP}. In these cases, we have only
annihilation or bremsstrahlung of the incident partons;
the resulting electromagnetic signal leaves the system unaffected by
any subsequent medium and its evolution. If such reactions show nuclear
effects, they are presumably due to initial state phenomena.
\par
After these more general remarks, we will now consider in detail the
use of quarkonium production as a probe for deconfinement in
dense strongly interacting matter. We concentrate
on quarkonium for several reasons.\ \J~suppression was predicted
\ref{Matsui} to be the consequence of QGP formation, and such a
suppression was subsequently indeed observed in high energy
nuclear collisions at the CERN-SPS \ref{NA38}. This
triggered an intensive study of possible alternative origins of such a
suppression. Hence the analysis necessary to establish an
unambiguous probe for deconfinement has been carried much further here
than for jets and can provide a good illustration of what needs to be
done before drawing any conclusions. In particular, as noted at the
beginning, we must understand theoretically and experimentally
the dynamics of the process to be used as probe and
\item{--}{how it is influenced by initial state nuclear effects,}
\item{--}{how it reacts to confined matter,}
\item{--}{how it reacts to deconfined matter, and}
\item{--}{how it reacts to non-equilibrium systems.}
\par\noindent
In section 2, we will therefore outline the theory of quarkonium
production and compare it to present data \ref{Quarko}. Section
3 will deal with quarkonium production in hadron-nucleus collisions.
In particular, we will see how the production of {\sl fast} quarkonia
provides us with information on the energy loss of a colour charge in
confined matter and on gluon shadowing. Next, in section 4, we will
discuss the conceptual basis of quarkonia as deconfinement probe and
present the results of the heavy quark theory for quarkonium-hadron
interactions. We then show how to test experimentally the resulting
transparency of confined matter to slow \J's. Section 5 will bring a
comprehensive comparison of the effect of confined and deconfined media
on quarkonium production; here we will recover the melting of the \J~in
a QGP \refs{Matsui}{Hwa} on a microscopic level. In Section 6, we
remove initial state nuclear modifications and study the effect of
pre-equilibrium deconfinement. Finally, in section 7, we give a a brief
summary and an assessment of what we have learned from present data.
\bigskip
\noindent
{\bf 2.\ Quarkonium Production in Hadron-Hadron Collisions}
\medskip
In this section, we shall sketch the basic dynamics of quarkonium
production in hadron-hadron collisions; for a summary, see
e.g.\ \ref{Quarko}. We shall speak about charmonium states; but
everything said also holds for bottonium.
\medskip\noindent
{\sl 2.1~~Colour Evaporation}
\medskip
The first stage of charmonium formation is the production of a $\c$
pair; because of the large quark mass, this process can
be described by perturbative QCD (Fig.\ 2.1). A parton from the
projectile interacts with one from the target; the parton
distributions within the hadrons are
determined e.g.\ by deep inelastic
lepton-hadron scattering. Initially, the $\c$ pair will in general be in
a colour octet state. It subsequently neutralises its colour and binds
to a physical resonance, such as \J, $\chi_c$ or \P. Colour
neutralisation
occurs by interaction with the surrounding colour field; this and the
subsequent resonance binding are presumably of non-perturbative nature
 (``colour evaporation" \ref{CE}). In the evaporation process,
the $\c$ pair can either combine with light quarks to form open
charm mesons ($D$ and $\bar D$) or bind with each other to form a
charmonium state.
\par
The basic quantity in this description is the total sub-threshold
charm cross section, obtained by integrating the perturbative $\c$
production over the mass interval from $2m_c$ to $2m_D$. At high
energy, the dominant part of \S~comes from gluon fusion (Fig.\ 2.1a), so
that we have
$$
\s(s) = \int_{2m_c}^{2m_D} d\hat s \int dx_1 dx_2~g_p(x_1)~g_t(x_2)~
\sigma(\hat s)~\delta(\hat s-x_1x_2s), \eqno(2.1)
$$
with $g_p(x)$ and $g_t(x)$ denoting the gluon densities in projectile
and target, respectively,
and $\sigma$ the $gg \to \c$ cross section. In pion-nucleon collisions,
there are also significant quark-antiquark contributions (Fig.\ 2.1b),
which become dominant at low energies.
The essential prediction of the colour evaporation model is
that the production cross section of any charmonium state $i$
is given by
$$
\sigma_i(s)~=~f_i~\s(s), \eqno(2.2)
$$
where $f_i$ is a constant which for the time being
has to be determined empirically. In other words, the energy
dependence of any charmonium production cross section is predicted to be
that of the perturbatively calculated sub-threshold charm cross section.
As a consequence, the production ratios of different charmonium
states
$$
{\sigma_i(s)\over \sigma_j(s)} = {f_i\over f_j} = {\rm const.}
\eqno(2.3)
$$
are predicted to be energy-independent. -- We note that
in the generalised colour evaporation model \ref{Quarko}, only a part of
the total subthreshold cross section $\s$ goes into charmonium
formation. In accord with perturbative open charm calcualtions, the
remainder (more than 50 \%) leads to $D{\bar D}$ production, with the
missing energy obtained by interaction with the colour field.
\bigskip
\vskip 6truecm
\centerline{Fig.\ 2.1: Production of a $c{\bar c}$
pair by gluon fusion (a) and $q{\bar q}$ annihilation (b)}
\bigskip\noindent
{\sl 2.2~~Quarkonium Production: Theory and Data}
\medskip
The predictions of the colour evaporation model have recently been
compared in a comprehensive survey \ref{Quarko} to
the available data, using parton distribution functions \refs{MRS}{GRV}
which take into account the new HERA results. In Figs.\ 2.2 and 2.3, we
see that the energy-dependence is well described for both \J~and
\U~production; for \J~production, the normalisation coeffficient is
$f_{\j}$=0.025. The \U~results are obtained for the sum of \U, \U' and
\U'' decaying into dimuons, with $Bf_{\u}=1.6\times
10^{-3}$ for the normalisation; here the branching ratios cannot be
directly removed. In the
fixed target/ISR energy range, the results from the two different sets
of parton distributions coincide; for the \J~at LHC energies, there is
some spread due to scale uncertainties in the parton distributions, which
hopefully can be removed by more precise DIS data. For the
\U~production, we have already now data up to 1.8 TeV, and in
Fig.\ 2.3 they are seen to agree very well with the prediction obtained
using the ``low energy" value $Bf_{\u}=1.6\times 10^{-3}$.
Previous phenomenological fits \ref{Craigie} had led to much smaller
rates; they are included (CR) in Figs.\ 2.2 and 2.3.
\par
In Figs.\ 2.4 and 2.5, the predicted energy-independence of
production ratios is found to hold as well,
again up to Tevatron energy. Here it should be noted that
the CDF data for the ratio \P/(\J) are taken at large transverse momenta
($5\leq p_T\leq 15$ GeV), while the lower energy data are integrated
over $p_T$, with the low $p_T$ region dominant. Hence colour evaporation
appears to proceed in the same way at both small and large $p_T$.
\par
The colour evaporation model does not determine the relative production
rates of the different states. In order to obtain them, the colour
evaporation process has to be specified in more detail. As an
example, we consider the ratios of the different $l=0$ states
shown in Figs.\ 2.4 and 2.5. Assume
that the initial colour octet state first neutralises its colour by
interaction with the surrounding colour field, producing a colour
singlet $\c$ state. The relative weights for \J~and \P~production can
then be expressed \ref{B-R} in terms of the corresponding masses and the
squared charmonium wave functions at the origin,
$$
{\sigma(\psi') \over \sigma(\psi)} = {R_{\psi'}^2(0) \over
R_{\psi}^2(0)}\left({M_{\psi}\over M_{\psi'}}\right)^5. \eqno(2.4)
$$
Here $\psi$ denotes the directly produced 1S $\c$ state, in contrast
to the experimentally observed \J, 40\% of which originates
from radiative $\chi_c$ decays \refs{Lemoigne}{Anton1}. The wave
functions
at the origin can in turn be related to the dilepton decay widths
$\Gamma_{ee} \sim (R^2(0) / M^2)$ \ref{Schuler}, giving
$$
{\sigma(\psi') \over \sigma(\psi)} = {\Gamma_{\psi'} \over
\Gamma_{\psi}}\left({M_{\psi}\over M_{\psi'}} \right)^3. \eqno(2.5)
$$
Inserting the measured values for masses and decay widths, we find
$$
{\sigma(\psi') \over \sigma(\psi)} \simeq 0.24~. \eqno(2.6)
$$
To compare this to the measured value of
$(\sigma(\psi')/\sigma(\psi))$,
we have to remove the $\chi_c$ contributions from the
experimental ratio,
$$
{\sigma(\psi') \over \sigma(\psi)} = \left[{1\over
1-(\sigma_{\chi}/\sigma_j)}\right] \left[
{\sigma(\psi') \over \sigma(\j)}\right]_{\rm exp}. \eqno(2.7)
$$
With the experimental values $\sigma(\psi')/\sigma(\j)\simeq 0.14$
(see Figs. 2.2 and 2.3) and $(\sigma_{\chi}/\sigma_j) \simeq 0.4$
\refs{Lemoigne}{Anton1},
this yields $\sigma(\psi')/ \sigma(\psi)\simeq 0.23$, in good
agreement with the theoretical result (2.6). We thus find that the
projection of the colour singlet $\c$ state onto \J~and \P~correctly
describes their production ratios at all energies and transverse
momenta.
\par
The predictions for direct bottonium production ratios corresponding to
Eq.\ (2.6) are
$$
{\sigma(\Upsilon') \over \sigma(\Upsilon)} \simeq 0.36~~;~~
{\sigma(\Upsilon'') \over \sigma(\Upsilon)} \simeq 0.27~. \eqno(2.8)
$$
Since the contributions from indirect production through
radiative $\chi_b$ decay are not yet known and there is also
feeddown from higher S-states, a quantitative comparison
is not possible here. Nevertheless, the predicted values differ by
less than 50 \% from the data and hence appear reasonable.
\bigskip\noindent
{\bf 3.\ Quarkonium Evolution and Hadron-Nucleus Collisions}
\medskip
Quarkonium production in a (finite) medium depends crucially on the
evolution stage in which the \Q~pair is during its passage. Here we
determine the different stages attainable in hadron-nucleus collisions
and discuss the effects which the nuclear environment has on them.
\medskip\noindent
{\sl 3.1~~Parton Fusion and Colour Neutralisation in Nuclear Matter}
\medskip
The first stage of charmonium production by hadronic interactions is, as
we saw, parton fusion resulting in a generally coloured $\c$ pair.
Subsequently, the $\c$ becomes colour neutral by emission or absorption
of gluons in the colour field of the interaction, and eventually this
small colour singlet state ``expands" to form a physical charmonium
resonance. If all or part of this evolution takes place inside a
nucleus, a number of effects can modify the production:
\item{--}{the effective distribution of the partons which fuse to form
the $\c$ can be modified by the nuclear environment (EMC effect);}
\item{--}{the colour octet $\c$ state will interact with the nuclear
medium strongly and without knowledge of its final charmonium state
(nuclear shadowing, $\c$ energy loss);}
\item{--}{the physical charmonium states will interact with the nuclear
medium according to their different cross sections on hadrons
(absorption).}
\par\noindent
Since the evolution is temporal, with finite time scales, fast
and slow charmonia (in the nuclear rest frame) will have quite different
fates. In this section, we shall determine the kinematic regimes for
the different evolution stages \refs{KS1}{KS5}; in the subsequent
sections, we will then discuss the dominant effects in each stage.
\par
To estimate how long the colour octet $\c$ state will live, we first
note that the \J~wave function keeps the charmed quarks close to their
mass shells; we thus have $p^2\simeq m_c^2$ (see Fig.\ 3.1 for
notation).
The intermediate quark with four-momentum $k=p+q$ is off-shell by an
amount
$$
\Delta = (p+q)^2-m_c^2 = 2pq,\eqno(3.1)
$$
where $q$ is the momentum of the third (colour neutralising) gluon;
in the spirit of the parton model, we keep all gluons on-shell.
In the low $p_T$ region, in which we are here primarily interested, this
third gluon can be arbitrarily soft; colour neutralisation could even
involve several gluons. In any case, the colour neutralisation process
cannot really be calculated in a purely perturbative framework. We shall
here nevertheless keep the structure of Fig.\ 3.1, taking it to be a
phenomenological extension into the non-perturbative soft-gluon regime.
A justification of such a procedure might come from recently developed
resummation methods \ref{Neubert}.
\bigskip
\vskip 6 truecm
\centerline{Fig.\ 3.1: \J~production through gluon fusion}
\bigskip
The proper life-time of the virtual coloured state is now given by the
uncertainty principle as
$$
\t_8 \simeq |\Delta|^{-1/2}. \eqno(3.2)
$$
In the rest frame of the \J, the life-time of the colour octet,
$$
\t_8\simeq (2m_c q_0)^{-1/2}, \eqno(3.3)
$$
is determined by the energy $q_0$ of the third gluon.
It is clear from Eqs.\ (3.1) and (3.2) that the colour octet state can
propagate over long distances only if the third gluon is soft enough.
In a confined medium, gluons cannot propagate over distances larger than
about 1 fm; hence the low energy cut-off is $q_0 \simeq \la \simeq 0.2$
GeV. This leads to
$$
\t_8 \simeq 0.25~{\rm fm} \eqno(3.4)
$$
for the colour neutralisation time. In the rest frame of the nuclear
target $A$, the $\c$ travels in this time a distance
$$
d_8 = \left( {P_A \over M}\right) \t_8, \eqno(3.5)
$$
where $P_A$ is the (lab) momentum of the $\c$ and $M$ its mass. From
Eq.\ (3.5) it is clear that in hadron-nucleus collisions sufficiently
fast $\c$ pairs will still be coloured when they leave the nuclear
medium.
\par
The average path length for a $\c$ produced in a $h-A$ collision is
$(3/4)R_A\simeq 0.86 A^{1/3}$ fm; for heavy nuclei, such as
$Pb~(A=208)$, this becomes about 5 fm. Hence charmonium states of lab
momenta $P_A~\gsim~60$ GeV have passed the nucleus as colour octets.
To relate this to the kinematic variables generally used in $h-A$
experiments, we
transform the lab momentum $P_A$ to the momentum $P$ measured in the
center of mass of a hadron-nucleon collision. The Feynman variable $x_F$
is then defined as $x_F\equiv P/P_{max}$, with $P_{max}(s)$ denoting the
maximum cms momentum possible for a charmonium state produced in a
hadron-nucleon collision of cms energy $\sqrt s$. From this, we obtain
\refs{KS1}{KS5} that for ${\sqrt s}~\gsim~20$ GeV and $x_F~\gsim~
0$, the measured charmonium states have traversed most of the nucleus as
colour octets. This kinematic range covers essentially all
high statistics charmonium production data taken in $h-A$ collisions
\refs{Badier}{E772}. These experiments therefore study the passage of a
colour octet $\c$ pair through nuclear matter; as such, they do not
provide any information on the interaction of fully formed physical
\J~or \P~with nucleons.
\par
An immediate consequence of this is that any nuclear effects observed
for charmonium production in $h-A$ collisions in the quoted kinematic
range must be the same for \J~and for \P~production. This conclusion
is indeed well satisfied experimentally (Fig.\ 3.2).
\par
We have so far considered the colour regime, i.e., the range in which
the $\c$ pair traverses the entire nucleus as a colour octet. When the
distance $d_0$ is less than 1 - 1.5 fm, the $\c$ is colourless when it
passes from the nucleon on which it was produced to the ``next"
nucleon inside the nucleus. At ${\sqrt s}\simeq 20$ GeV, this occurs for
$x_F~\lsim~-0.2$. At this point, the $\c$ is still a generic small
colour singlet state and not yet a specific physical resonance.
The other extreme to the colour regime is the range in
which the $\c$ has become a fully formed physical resonance (\J, \P)
before it leaves the environment of the nucleon at which it was formed.
Since the different states have different sizes and hence different
formation times, the resonance regime will be state-specific, i.e., at
a given collision energy, it will cover a larger $x_F$-region
for the \J~than for the \P. The distance $d_r(i)$ travelled by the
$\c$ before becoming a physical resonance state $i$ is given by
$$
d_r(i) = \left( {P_A \over M} \right) \t_r(i) \eqno(3.6)
$$
where $\t_r(i)$ denotes the resonance formation time. Potential theory
\ref{Karsch} leads to the estimates $t_r(\j) \simeq 0.35$ fm and
$t_r(\p)\simeq1$ fm. For a collision energy ${\sqrt s} \simeq 20$ GeV,
this means that a \J~has become a fully formed resonance for
$x_F~\lsim -0.5$. The \P~is still nascent even for $x_F=-1$; although
it is a colour singlet on its passage through the nucleus if
$x_F ~\lsim~-0.2$, it has not yet reached its full physical size when it
leaves the nucleus even for $x_F=-1$.
\par
In the transition regime between these two extremes, the $\c$
experiences colour interactions on part of its passage, while it
traverses the medium as a small colour singlet on the remaining part.
In Table 1, we have summarised the different kinematic ranges for
proton beam incident on a $Pb$ target at a cms nucleon-nucleon
collision energy ${\sqrt s}=20$ GeV, for (1S) \J~and (2S) \P~production.
In the following sections, we survey the possible nuclear effects which
can arise in the different kinematic regimes.
\par
$$
{\offinterlineskip \tabskip=0pt
\vbox{
\halign to 0.9\hsize
{\strut
\vrule width0.8pt\quad#
\tabskip=0pt plus100pt
& # \quad
&\vrule#&
&\quad \hfil # \quad
\tabskip=0pt
&\vrule width0.8pt#
\cr
\noalign{\hrule}\noalign{\hrule}
&~~~~~~~~~ &&~~~~~~~~~~~~~&&~~          ~~~~~~~&\cr
& Regime   &&     \J      &&       \P          &\cr
&~~~~~~    &&~~~~~~~~~~~~~&&              ~~~~~&\cr
\noalign{\hrule}
&~~        &&~~           &&                   &\cr
& Colour Regime  && $x_F \geq 0$  &&  $x_F \geq 0$     &\cr
&~~        &&~~             &&~                  &\cr
\noalign{\hrule}
&          &&               &&                   &\cr
& Transition Regime
           &&  $-0.5 \leq x_F \leq 0$  && $-1 \leq x_F \leq 0$ &\cr
&~~        &&~~             &&~                  &\cr
\noalign{\hrule}
&~~        &&               &&                   &\cr
& Resonance Regime   &&  $-1 \leq x_F \leq -0.5$ &&  ---
&\cr
&          &&             &&                     &\cr
\noalign{\hrule}}
}}
$$
\par
\centerline{Table 1:}
\par
\centerline{Charmonium Production in $p-A$ Collisions at
$\sqrt{s}$ = 20 GeV}
\bigskip\noindent
\medskip\noindent
{\sl 3.2~~Quantum-Mechanical Interference and Nuclear Shadowing}
\medskip
We now address nuclear modifications of charmonium production
due to what at first sight appears to be a change in the effective
parton distribution function $g_t(x_t)$
in nuclei, compared to that in nucleons. The fraction of the target
momentum carried by the corresponding parton in the elementary
interactions shown in  Fig.\ 2.1 is given by
$$
x_t = {1\over 2}(\sqrt{(4M^2/s)+x_F^2} \pm x_F), \eqno(3.7)
$$
where $M$ is the mass of the $\c$ system; the $\pm$ signs hold for
$x_F>0$ and $x_F<0$, respectively. For $\c$ production at $x_F\geq 0$
and $\sqrt s \geq 20$, $x_t \leq 0.15$. In this $x_t$ region,
deep inelastic scattering experiments on nuclei find a lower quark
density than on nucleons (nuclear shadowing \ref{Shadow}),
as illustrated in Fig.\ 3.3 (the solid line is phenomenological
fit \ref{Smirnov}.
Such nuclear shadowing can, however, not be interpreted as
an intrinsic change of the parton distribution in nuclei. As
such, it would be applicable in factorisable form
to all hard processes; but for Drell-Yan dilepton production by
$p-A$ collisions, which involves nuclear quark distribution in the same
$x_t$ region, little or no modification is observed \ref{E772}. The
appearent puzzle is resolved by noting that nuclear shadowing
is due to quantum-mechanical interference, similar to
the well-known Landau-Pomeranchuk effect \ref{LP}. This interference
depends on the specific process and hence leads to different nuclear
modifications for charmonium hadro- and photoproduction, for Drell-Yan
dilepton production and for deep inelastic lepton-hadron scattering
\ref{KS2}. For a first discussion of quantum interference effects
in deep inelastic scattering, see \ref{Lu}.
\par
We had seen above in Eq.\ (3.5) that sufficiently fast $\c$ pairs
traverse the entire nucleus in a virtual coloured state. In this case,
the interactions of the pair with different nucleons cannot be added
incoherently and factorisation breaks down even for dynamically
uncorrelated nucleons: quantum mechanical interference can now
lead to nuclear modifications even though both the parton
distribution and the elementary parton interaction amplitude are
unchanged. Such interference effects set in when the coherence length
$d_c=(1/mx_t)$, over which the $\c$ is in an off-shell
virtual state, becomes larger than
the internucleonic distance $d_0=n_0^{-1/3}\simeq 1.8$ fm, with
$n_0=0.17$ fm$^{-3}$ denoting standard nuclear density and $m$
the nucleon mass. This distance
has to be compared to the mean free path
$\l_8$ of the virtual colour state in nuclear matter,
$$
\l_8 = (\sigma_8 n_0)^{-1}, \eqno(3.8)
$$
where  $\sigma_8$ is the cross section for the interaction of the colour
octet $\c$ with nucleons. To estimate the size of $\sigma_8$, we
assume that the
overall colour neutrality required for the propagation of the
coloured $\c$ through the nucleus is provided by a comoving light $\q$
pair. The interaction cross section of the \Q$\q$ system is then defined
by the light $\q$ pair, so that its size can be as large as that of
light $\q$ mesons (20 - 30 mb). This is much larger than
that of a physical \J; it is now not the size of the $\c$, but its
colour which determines the interaction. The mean free path
corresponding
to such a cross section becomes $\l_8 \simeq 2 - 3$ fm, which is about
the size of the internucleonic distance $d_0=n_0^{-1/3}\simeq 1.8$ fm,
so that charmonium production
in $p-A$ collisions for $x_F\geq 0$ and $\sqrt s \geq 20$ GeV satisfies
the ``shadowing relation" $d_0 \simeq \l_8 \ll d_c$ \ref{KS2}. In this
regime, the interactions of the virtual $\c$ with successive nucleons
interfere destructively with each other; it is the regime which for
electromagnetic interactions results in the Landau-Pomeranchuk effect.
The production cross section $\sigma(pA \to \c)$ on nuclei is here
therefore less than the incoherent result $A \sigma(pp \to \c)$.
\par
The basic quantity studied experimentally is the ratio
$$
R_{A/p} = {\sigma(pA \to \c) \over A\sigma(pp \to \c)}; \eqno(3.9)
$$
it would be unity in the absence of any nuclear effects. It is observed,
however, that $R_{A/p}$ decreases with
increasing $\c$ momentum in the nuclear rest frame (i.e., whenever
$d_c$ increases), as well as with increasing $A$.
Rather than in terms of $d_c$, $R_{A/p}$ is generally
studied in terms of
the fractional target parton momentum $x_t=(md_c)^{-1}$.
Decreasing $x_t$ thus increases
the path length for the colour octet inside the nucleus, and a longer
path of the virtual state leads to more destructive interference.
When $d_c$ reaches the nuclear diameter $2R_A$, we have the maximum
possible interference, so that the suppression saturates
for $x_t \leq x_t^{\rm sat} \simeq (2mR_A)^{-1}$, with the
approximate limit \ref{KS2}
$$
R_{A/p}^{sh}(x_t^{\rm sat}) \simeq {\l_8 \over R_A}.
\eqno(3.10)
$$
A more realistic form including geometric effects is given by
$$
R_{A/p}^{sh}(x_t^{\rm sat}) \simeq 1 - e^{-\l_8/R_A}, \eqno(3.11)
$$
which reduces to Eq.\ (3.10) for $\l/R_A \ll 1$. The available data fall
into the region $0 \leq R_A/\l_8 = n_0 \sigma_8 R_A~\lsim~0.7$, so that
the measured nuclear reduction lies between 0.5 and 1. In this region,
the relation $e^{-x} \simeq 1 - e^{1/x}$ is quite well satisfied, so
that
$$
R_{A/p}^{sh}(x_t^{\rm sat}) \simeq  e^{-n_0 \sigma_8 R_A} \eqno(3.12)
$$
also should also describe the data. The form (3.12) was in fact obtained
as an empirical fit \ref{Gerschel}; we see here, however, that it is the
result of quantum-mechanical interference effects of the virtual $\c$
and is not related to the interaction of a physical \J~with nucleons, as
assumed there.
\par
 For $d_c \leq d_0$, the scattering on different
nucleons becomes incoherent, giving $R_{A/p}^{sh}=1$,
with $x_t \geq x_t^{\rm inc} \simeq (md_0)^{-1}$. For $A=200$, the two
limits are $x_t^{\rm sat}=0.015$ and $x_t^{\rm inc}=0.12$.
Qualitatively, this is the type of behaviour
found in charmonium production for $x_F$ not too large \ref{GS1}.
The specific functional form of the suppression between the two
extremes can only be estimated phenomenologically. Experimental
results have often been parametrised as $R_{A/p} \simeq A^{\alpha - 1}$;
our interference considerations imply that the coefficient $\alpha$
depends on $x_t$. An analysis of charmonium production in $p-A$
collisions \ref{GS1} suggests a simple linear fit in $\ln A$ and
$\ln x_t$,
$$
R_{A/p}^{sh} \simeq [1 - c~\ln (x_t md_0) \ln A]. \eqno(3.13)
$$
The scale $md_0 \simeq 0.12$ is chosen to make $R_{A/p}^{sh}=1$ in
the incoherence regime; the constant $c$ has to be chosen such as to
give the correction saturation value. For a fit of $p-A$ data to
a similar form, see \ref{KS1}.
\medskip\noindent
{\sl 3.3~~Momentum Loss of a Colour Charge in Nuclear Matter}
\medskip
The suppression $R_{A/p}^{sh}$ cannot be directly compared to data,
however,
since shadowing is not the only effect suffered by the coloured $\c$
state on
its path through the nucleus. The $\c$ is formed in the collision of the
projectile with one of the target nucleons. To leave the nuclear medium,
it has to traverse the remaining part of the nucleus, in which it can
interact with other nucleons. Getting from one nucleon to the next means
passing regions of internuclear physical vacuum, which does not support
colour charges. To estimate the effect of this passage, we imagine the
$\c$ to stretch a string from its formation region to the next
nearest nucleon, then from this to the one after that, and so on. If the
pair intially had a cms momentum $P$, then this will be shifted to
$$
P' = P - \kappa_8(L_A-d_0), \eqno(3.14)
$$
where $\kappa_8 \simeq (9/4) \kappa \simeq (9/4)$ GeV/fm is the string
tension of the colour octet, $\kappa \simeq $ 1 GeV/fm that of the
fundamental triplet \ref{KS1}; $L_A$ is the total path length of the
colour state. As consequence of this momentum loss, a
$\c$ pair observed at a given $x_F$ must have been originally produced
at a higher value $x_F/\delta $, with
$$
\delta = \left( {P-\kappa_8(L_A-d_0) \over P} \right). \eqno(3.15)
$$
The resulting normalised $x_F$-distribution $F_A(x_F)
\equiv(1/A)(d\sigma_{hA\to \j}/dx_F)$ for \J~production in $h-A$
collisions thus becomes
$$
F_A(x_F)~=~W_0 F_0(x_F) +
(1-W_0)\left({F_0(x_F/\alpha)\over\delta}\right)\Theta(1-x_F/\delta).
\eqno(3.16)
$$
Here $F_0(x_F)$ is the $x_F$-distribution in $h-p$ collisions and
$$
W_0\simeq\exp\{-(L_A-z_0)/d_8\} \eqno(3.17)
$$
gives the probability that the $\c$ pair
will not undergo any scattering on its way out of the target;
thus $W_0=1$
implies no nuclear \J~suppression. The second term in Eq.\ (3.16)
gives the effect of the $n$ coherent scatterings in the medium, with the
resulting shift in $x_F$. The factor $1/\delta$ assures that the
distribution remains normalised. Note that the momentum loss of $\c$
pair, for $x_F\geq 0$ as described here, does not imply any integrated
\J~suppresion: the production is simply shifted from larger to smaller
$x_F$.
Dividing Eq.\ (3.16) by the $x_F$-distribution for h-p collisions,
$F_0(x_F)$, gives us the ratio of $h-A$ to $h-p$ distributions,
$$
R_{A/p}^{ml}(x_F)~=~W_0 + (1-W_0)\left\{{F_0(x_F/\alpha)\over
\alpha F_0(x_F)}\right\}\Theta(1-x_F/\alpha). \eqno(3.18).
$$
For a nucleus of $A=200$ and the other parameters as above,
we get the dependence on $x_F$ shown in Fig.\ 3.4 for three
different beam energies. We note that on nuclei, compared to nucleons
as target, \J~production is shifted to lower $x_F$, essentially because
of the momentum degradation of the colour octet in passing through the
medium. This momentum loss saturates at
$\Delta P_{\rm max} \simeq \kappa_8(R_A-z_0)\simeq 10$ GeV and is thus
bounded. As a consequence, the relative shift in $x_F$ decreases with
increasing beam energy, and the effect of the medium disappears
for $\sqrt s \to \infty$, when $R_{A/p}^{ml}\simeq 1$ for $0<x_F<1$.
\bigskip
\vskip 7 truecm
\noindent Fig.\ 3.4: The effect of colour octet interactions on
charmonium production in $p-A$ collisions at three different energies
\bigskip
In addition to the non-perturbative nuclear modification of fast
quarkonium production in nuclei, the $\c$ pair can undergo hard
interactions with the quarks and gluons within each of the nucleons it
passes through. We neglect this here, since the density of sufficiently
hard scatterings in nuclei is too low to compete with string stretching,
as will become evident in Section 3.4. This is changed, however, in a
quark-gluon plasma, and the difference in parton hardness for confined
and deconfined media is in fact crucial for the use of quarkonia as
confinement/deconfinement probe.
\par
The observed suppression of quarkonium production in $h-A$ collisions
will be due to both the effects just discussed, nuclear shadowing and
momentum loss of the coloured $\c$ pair. We thus have to compare
experimental results to the product
$$
R_{A/p}(x_F) = R_{A/p}^{sh}(x_F) \times R_{A/p}^{ml}(x_F) \eqno(3.19)
$$
of the two mechanisms, as given by Eqs.\ (3.12) and (3.18).
The energy loss is effective mainly at larger $x_F$; since the
integrated cross section, on the other hand, is determined by the
region around $x_F \simeq 0$, its suppression is dominantly due to
nuclear shadowing. Hence Eq.\ (3.11) should determine the main
$A$-dependence of charmonium suppression in $p-A$ collisions at $x_f
\geq 0$, and as a comparison to the equivalent form (3.12) shows
\ref{Gerschel}, it indeed does.
\par
This form, but
with a slightly different parametrisation
of shadowing \ref{GS1}, has been compared to the available data
\refs{Badier}{E772} in \ref{KS1}. In spite of the simplistic
fit to shadowing and some kinematic approximations at small
$x_F$, the result (Fig.\ 3.5) is seen to reproduce the observed
suppression quite well.
\par
Quarkonium production in hadron-nucleus collisions, as experimentally
studied up to now, appears thus as theoretically quite well
understood, based on the interaction of a fast colour octet
$\c$ pair with nuclear matter. The dominant mechanisms for this
interaction are the destructive quantum-mechanical interference of the
scattering amplitudes of the $\c$ on different nucleons, and the
momentum loss of the $\c$ as it traverses the physical vacuum between
successive nucleon interactions.
\medskip\noindent
{\sl 3.4~~Charmonium Interactions in Nuclear Matter}
\medskip
If we want to know the effect of nuclear matter on a fully formed
physical \J, then -- as shown in Table 1 -- we have to study the
production of \J's slow in the nuclear rest frame.
The study of such slow charmonia has up to now been essentially
impossible. It requires the detection of slow decay dileptons, and for
this the abundance of slow
hadrons constitutes an overwhelming background. In the case of
fast dileptons, a hadron absorber can eliminate these, and hence all
$p-A$ studies were so far restricted to dilepton pairs of more than 20
GeV in the rest frame of the nuclear target.
The advent of the $Pb$-beam at the CERN-SPS has removed this constraint.
With the $Pb$-beam incident on a hydrogen (or deuterium) target, the
nuclear rest frame moves with a lab momentum of 160 GeV.
Hence now those charmonia (and their decay dileptons) which are slow in
the nuclear rest frame are very fast in the
lab system and will thus pass the hadron absorber; such experiments can
therefore provide the cross section for the break-up interaction of
physical \J's on nucleons. To avoid confusion, we shall
continue to define $x_F$ also in such $Pb-h$ reactions as positive for
cms momenta in the direction of the hadron; with this terminology,
$Pb-h$ collisions provide us with information about the so far unknown
region of negative $x_F$. Estimates \refs{KS1}{KS5} show that for
$\sqrt s \geq 17$ GeV and $x_F \leq -0.2$, a $\c$ is colourless before
it reaches the next nucleon, so that colour interactions no longer
enter.
\par
We now parametrise the survival probability for state $i$
($i=\j,~\chi_c,~\p$) as
$$
S_i = exp\{-n_0 \sigma_i L\}, \eqno(3.20)
$$
where $n_0$ is as before normal nuclear density,
$L$ the average path length of the charmomium state $i$ in the nucleus,
and $\sigma_i$ its absorption cross section of the state $i$ in
nuclear matter. If the charmonium state is fully formed before it
leaves the
range of the nucleon at which it was produced, $\sigma_i$ is simply
the charmonium-nucleon cross section. If it is not yet fully formed, the
effective cross section will be smaller, vanishing in the colour
transparency limit of a pointlike colour singlet
\ref{colour}. For illustration, we can therefore use a
simplistic parametrisation of the
cross section as function of the distance $d$ the state has travelled
before leaving the nucleus \refs{Farrar}{Blaizot},
$$
\sigma_i(d) = \sigma_i \left( {d\over {\bar L_i}}\right)^2, \eqno(3.21)
$$
where $\sigma_i$ is the fully developed cross section and ${\bar
L}=[(P_A/M_i)\t_r^{(i)} - 1]$ the effective distance travelled until
full resonance
formation. Eq.\ (3.21) holds for $d\leq {\bar L}$; for $d\geq {\bar
L}$,
$\sigma_i(d)=\sigma_i$. Using this parametrisation, Eq.\ (3.20) is
replaced by
$$
S_i = exp\{-n_0\sigma_i[L - {2\over 3} {\bar L_i}]\}; \eqno(3.22)
$$
this reduces to Eq.\ (3.20) for momenta high enough to make ${\bar
L}=0$. To determine the actual survival
probabilities, we now need the values $\sigma_i$ for the fully
formed resonances colliding with nucleons. That is a topic in its own
right and will be taken up in the next chapter. Before turning to this
question, we want to connect the survival probability (3.20) to
the measured suppression ratio $R_{A/p}$.
\medskip\noindent
{\sl 3.5 EMC Effect Modifications}
\medskip
The survival probabilities $S_i$ are related to the
$p-A$ and $p-p$ induced charmonium production cross sections through
$$
{(d\sigma_i^{pA}/dx_F)\over A(d\sigma_i^{pp}/dx_F)} = {g_A(x)\over
g_p(x)} S_i(x_F) \equiv R_{A/p}(x) S_i(x_F), \eqno(3.23)
$$
where $g_A(x)$ and $g_A(x)$ are the parton distribution functions in
nuclear and proton target, respectively. The fractional parton momentum
$x$ is now, with $x_F \leq 0$, expressed by
$$
x = {1\over 2}(\sqrt{(x_F^2 + (4M_i^2/s))} - x_F) \eqno(3.24)
$$
in terms of the variables $x_F$ and $s$. At the energy possible for
$Pb$-beam experiments (${\sqrt s}=17.4$ GeV), and for $x_F\leq 0$, we
have  $x\geq 0.15$. We are thus above the region in which
quantum-mechanical coherence effects (nuclear shadowing) can play a
role; the reaction now is indeed an incoherent sum of interactions
with different nucleons. In deep inelastic scattering on nuclear
targets there is for $x_2\geq 0.15$ a modification
in comparison to the same process on nucleons (EMC effect \ref{EMC};
it can now be interpreted as a genuine change of the parton
distribution function and applied to other hard processes. The resulting
pattern for quark distributions $q(x)$,
$$
R^{EMC}_{A/p}(x_F) = {q_A(x_F) \over q_p(x_F)} \eqno(3.25)
$$
is illustrated in Fig.\ 3.6 for $M=3.1$ GeV and $\sqrt s = 17.4$ GeV,
making use of Eq.\ (3.24) to convert the Bjorken variable $x$ to $x_F$.
\par
Although the EMC effect has so far been observed only for quarks, it is
to be expected that gluons will exhibit a similar behaviour. This is one
reason why the initial state factor $R_{A/p}(x)$ will introduce an
further $x_F$ variation in addition to that coming from $S_i(x_F)$. A
second reason was already indirectly mentioned above: with increasing
$|x_F|$, charmonium production is more and more due to
quark-antiquark annihilation rather than to gluon fusion.
The two contributions become approximately equal
around $|x_F|=0.5$, and as $|x_F| \to 1$, the $q{\bar q}$ contribution
is dominant. We shall here assume that
the gluon distributions behave similarly and use the quark form of
$R_{A/p}$ in the whole region $-1.0 \leq x_F\leq 0$. Eqs. (3.22)/(3.23)
and the $R^{EMC}_{A/p}$ of Fig.\ (3.6) then predict the behaviour of
the measured cross section ratio in the region of negative $x_F$.
\bigskip\noindent
{\bf 4.\ The Theory of Quarkonium-Hadron Interactions}
\medskip
In this section, the cross sections for the inelastic interaction
between quarkonia and light hadrons will be calculated in short distance
QCD, based on the small radii and the large binding energies of the
lowest \Q~resonances. A break-up of such states requires hard gluons,
but the gluons confined to hadrons slow in the quarkonium rest frame
are in general very soft, so that the resulting cross sections remain
very small until quite high collision energies.
\refadd{Peskin}
\refadd{Bhanot}
\refadd{SVZ}
\refadd{Novikov}
\refadd{Kaidalov}
\medskip\noindent
{\sl 4.1~~The Short-Distance Analysis of Quarkonium-Hadron Interactions}
\medskip
The interaction of quarkonia
with ordinary light hadrons plays an important role both in the
dynamics and in the thermodynamics of strong interaction physics.
For QCD dynamics, it is important since the small quarkonium size probes
the short-distance aspects of the big light hadrons and thus makes
a parton-based calculation of the overall cross section possible
\refto{Peskin}{Kaidalov}. In QCD thermodynamics,
quarkonia can be used as a probe for deconfinement \refs{Matsui}{KS3},
provided their interaction in dense confined matter
can be distinguished from that in a quark-gluon plasma.
\par
We begin this chapter with a brief summary of the QCD analysis of
quarkonium interactions with light hadrons. It concludes that the
small size of quarkonia, combined with the rather large mass gap to
open charm or beauty, strongly inhibits their break-up by low energy
collisions with light hadrons \ref{Bhanot}; the total quarkonium-hadron
cross section attains a constant asymptotic value
only at very high energies, compared to corresponding cross sections for
the interaction between light hadrons. This slowly rising form of the
cross section is derived from an operator product
expansion with ensuing sum rules and becomes quite transparent in parton
language.
\par
The QCD analysis of quarkonium interactions applies
to heavy and strongly bound quark-antiquark
states \ref{Bhanot}; therefore we here restrict ourselves to the
lowest $\c$ and $\b$ vector states \J~and \U, which we denote
generically by \F, following the notation of \ref{Bhanot}.
For such states, both the masses $m_Q$ of the constituent
quarks and the binding energies $\e_0(\f) \simeq (2M_{(Qq)}
- M_{\f}$) are much larger
than the typical scale $\Lambda_{\rm QCD}$ for non-perturbative
interactions; here $(Qq)$ denotes the lowest open charm or beauty
state. In $\f-h$ interactions, as well as in \F-photoproduction,
$\gamma h \to \f h$, we thus only probe a
small spatial region of the light hadron $h$; these processes are much
like deep-inelastic lepton-hadron scattering, with large $m_Q$ and
$\e_0$ in place of the large virtual photon mass $\sqrt{-q^2}$.
As a result, the calculation of \F-photoproduction and of
absorptive $\f-h$ interactions
can be carried out in the short-distance formalism
of QCD. Just like deep-inelastic leptoproduction, these reactions
probe the parton structure
of the light hadron, and so the corresponding cross sections can
be calculated in terms of parton interactions and structure
functions.
\par
In the following, we shall first sketch the theoretical basis which
allows quarkonium interactions with light hadrons to be treated
by the same techniques as used in deep-inelastic lepton-hadron
scattering or in the photoproduction of charm. We show the derivation
for the sum rules which relate the absorptive $\f-h$ cross section
to hadronic gluon structure functions \refs{Peskin}{Bhanot}.
This relation given, we calculate explicitly the energy dependence of
the cross section. Readers only interested in this
behaviour can therefore go immediately to Eq.\ (4.24).
\par
Consider the amplitude for forward scattering of a virtual
photon on a nucleon,
$$
F(s, q^2) \sim
i\int d^4x e^{iqx} <N|T\{J_{\mu}(x)J_{\nu}(0)\}|N>. \eqno(4.1)
$$
In the now standard application of QCD to
deep-inelastic scattering one exploits the fact that at large
spacelike photon momenta $q$ the amplitude is dominated by small
distances of order $1/\sqrt{-q^2}$ (Fig. 4.1a).
The Wilson operator product expansion then
allows the evaluation of the amplitude at the unphysical point
$pq\to 0$, where $p$ is the four-momentum of the nucleon.
Since the imaginary part of the amplitude (4.1) is proportional
to the experimentally observed structure functions of deep-inelastic
scattering, the use of dispersion relations relates the value
of the amplitude at $pq\to 0$ point to the integrals over the
structure functions, leading to a set of dispersion sum rules
\ref{Gross}.
The parton model can be considered then as a particularly useful
approach satisfying these sum rules.
\par
In the case $J_{\mu} = \bar{Q} \gamma_{\mu} Q$, i.e., when
vector electromagnetic current in Eq.\ (4.1) is that
of a heavy quark-antiquark pair, large momenta $q$ are not needed
to justify the use of perturbative methods.
Even if $q\sim 0$, the small space-time scale of $x$ is set by the
mass of the charmed quark, and the characteristic distances
which are important in the correlator (4.1) are of the order of
$1 /2m_Q$ (Fig.\ 4.1b). In \refs{SVZ}{Novikov}, this observation was
used to
derive sum rules for charm photoproduction in a manner quite similar to
that used for deep-inelastic scattering.
\par
In the interaction of quarkonium with light hadrons, again the
small space scale is set by the
mass of the heavy quark, and the characteristic distances involved
are of the order of quarkonium size, i.e., smaller than the
non-perturbative hadronic scale $\Lambda_{QCD}^{-1}$.
Moreover, since heavy quarkonium and light hadrons do not have
quarks in common, the only allowed exchanges are purely gluonic.
However, the smallness of spatial size is not enough to justify the use
of perturbative expansion \ref{Bhanot}. Unlike in the case of
\F-photoproduction, heavy quark lines now appear in the initial
and final states (see Fig.\ 4.1c), so that the \Q~state can emit and
absorb gluons at points along its world line widely separated in
time. These gluons must be hard enough to interact
with a compact colour singlet state (colour screening leads to a
decoupling of soft gluons with the wavelengths larger than the
size of the \F); however, the interactions among the gluons
can be soft and nonperturbative. We thus have to assure that the
process is compact also in time.
Since the absorption or emission of a gluon turns
a colour singlet quarkonium state into a colour octet,
the scale which regularizes the time correlation of such processes
is by the quantum-mechanical uncertainty
principle just the mass difference between the colour-octet and colour-
singlet states of quarkonium: $\tau_{c} \sim 1/ (\e_8 - \e_1)$.
The perturbative Coulomb-like piece of the heavy quark-antiquark
interaction
$$
V_k(r)=-g^2 {c_k \over {4\pi r}} \eqno(4.2)
$$
is attractive in the colour singlet ($k=1$) and repulsive in the
colour-octet ($k=8$) state; in SU(N) gauge theory
$$
V_1=-{g^2 \over {8\pi r}} {N^2-1 \over N}, \eqno(4.3a)
$$
$$
V_8={g^2 \over {8\pi r}} {1 \over N}. \eqno(4.3b)
$$
To leading order in $1/N$, the mass gap between the singlet and octet
states is therefore just the binding energy of the heavy
quarkonium $\e_0$, and the characteristic correlation time for
gluon absorption and emission is
$$
\tau_c \sim 1/\e_0. \eqno(4.4)
$$
Although the charm quark is not heavy enough to ensure a pure Coulomb
regime even for the lowest $c\bar{c}$ bound states ($\eta_c$ and $\j$),
the mass gap determined from the observed value of open charm threshold
clearly shows that $\tau_c < \Lambda_{QCD}$. For the \U, the interaction
is in fact essentially Coulomb-like and the mass gap to open beauty
is even larger than for charm. One therefore expects to be able to treat
quarkonium interactions with light hadrons by the same QCD methods
that are used in deep-inelastic scattering and charm photoproduction.
\par
We thus use the operator product
expansion to compute the amplitude of heavy quarkonium
interaction with light hadrons,
$$
F_{\f h} = i\int d^4x e^{iqx} \langle h|T\{J(x)J(0)\}|h \rangle =
\sum_n c_n(Q,m_Q) \langle O_n \rangle ,
\eqno(4.5)
$$
where the set $\{O_n\}$ includes all local gauge invariant operators
expressible in terms of gluon fields; the matrix elements $\langle
O_n \rangle$ are taken between the initial and final light-hadron
states. The coefficients $c_n$ are computable perturbatively
\ref{Peskin} and process-independent. As noted above, in
deep-inelastic scattering
the expansion (4.5) is useful only in the vicinity of the point
$pq\to 0$. The same is true for the case of quarkonium interaction
with light hadrons. As shown in \ref{Bhanot}, the expansion (4.5) can
therefore be rewritten as an expansion in the variable
$$
\lambda = {pq \over M_{\f}} = {(s-M_{\f}^2 - M_h^2)\over 2M_{\f}}
\simeq {(s-M_{\f}^2)\over 2M_{\f}}
\eqno(4.6)
$$
where $M_h$ is the mass of the light hadron;
the approximate equality becomes valid the heavy quark limit.
For the lowest 1S quarkonium state one then obtains
$$
F_{\f h} = r_0^3\ \e_0^2\ \sum_{n=2}^{\infty}
d_n \langle O_n \rangle
\left({\lambda \over \e_0}\right)^n, \eqno(4.7)
$$
where $r_0$ and $\e_0$ are Bohr radius and binding energy of the
quarkonium, and the sum runs over even values of $n$ to ensure the
crossing symmetry of the amplitude.
The most important coefficients $d_n$ were computed
in \ref{Peskin} to leading order in $g^2$ and $1/N$.
\par
Since the total $\f-h$ cross section $\sigma_{\f h}$ is proportional
to the imaginary part of the amplitude $F_{\f h}$, the dispersion
integral over $\l$ leads to the sum rules
$$
{2\over \pi}\int^{\infty}_{\l_0} d\l~\l^{-n} \sigma_{\f h}(\l)
= r_0^3~ \e_0^2~ d_n \langle O_n \rangle \left( {1\over \e_0}
\right)^n.
\eqno(4.8)
$$
Eq.\ (4.7) provides only the inelastic intermediate states in the
unitarity
relation, since direct elastic scattering leads to contributions of
order $r_0^6$. Hence the total cross section in Eq.\ (4.8) is due to
absorptive interactions only \ref{Bhanot}, and the integration
in Eq.\ (8) starts at a lower limit $\l_0 > M_h$.
 Recalling now the expressions for radius and binding energy of
1S Coulomb bound states of a heavy quark-antiquark pair,
$$
r_0 = \left( {16\pi \over {3 g^2}} \right) {1 \over m_Q}, \eqno(4.9)
$$
$$
\e_0 = \left( {3 g^2 \over {16 \pi}} \right)^2 m_Q, \eqno(4.10)
$$
and using the coefficients $d_n$ from
\ref{Peskin}, it is possible \ref{Bhanot} to rewrite these sum rules
in the form
$$
\int_{\l_0}^{\infty} { d\lambda \over \l_0} \left( {\lambda \over
\l_0} \right)^{-n}
\sigma_{\f h} (\lambda) = 2 \pi^{3/2}\ \left(16 \over 3 \right)^2\
{{\Gamma \left(n + {5\over 2} \right)} \over {\Gamma (n + 5)}}\
\left( {16\pi \over {3g^2}}\right)\
{1 \over m_Q^2}\ \langle O_n \rangle, \eqno(4.11)
$$
with $\l_0/\e_0 \simeq 1$ in the heavy quark limit.
The contents of these sum rules become more transparent
in terms of the parton model. In parton language,
the expectation values $\langle O_n \rangle$ of the operators composed
of gluon fields can be expressed as Mellin
transforms \ref{Parisi} of the gluon structure function of the light
hadron, evaluated at the scale $Q^2=\e_0^2$,
$$
\langle O_n \rangle = \int_0^1 dx\ x^{n-2} g(x, Q^2 = \e_0^2).
\eqno(4.12)
$$
Defining now
$$
y = {\l_0 \over \lambda} , \eqno(4.13)
$$
we can reformulate Eq.\ (11) to obtain
$$
\int_0^1 dy\ y^{n-2} \sigma_{\f h}(\l_0 / y) = I(n)\
\int_0^1 dx\ x^{n-2} g(x, Q^2 = \e_0^2), \eqno(4.14)
$$
with $I(n)$ given by
$$
I(n) =  2 \pi^{3/2}\ \left(16 \over 3 \right)^2\
{{\Gamma \left(n + {5\over 2} \right)} \over {\Gamma (n + 5)}}\
\left( {16\pi \over {3g^2}}\right)\ {1 \over m_Q^2}. \eqno(4.15)
$$
Eq.\ (4.14) relates the $\f-h$ cross section to the gluon structure
function. To get a first idea of this relation, we neglect the
$n$-dependence of $I(n)$ compared to that of $\langle O_n \rangle$; then
we conclude that
$$
\sigma_{\f h}(\l_0 / x) \sim g(x, Q^2=\e_0^2), \eqno(4.16)
$$
since all order Mellin transforms of these quantities are equal up to a
constant. From Eq.\ (4.16) it is clear that
the energy dependence of the $\f-h$ cross section is entirely
determined by the $x-$dependence of the gluon structure function.
The small $x$ behaviour of the structure function governs
the high energy form of the cross section, and the hard tail
of the gluon structure function for $x \to 1$ determines
the energy dependence of $\sigma_{\f h}$ close to the
threshold.
\par
To obtain relation (4.16), we have neglected the $n$-dependence of the
function $I(n)$. Let us now try to find a more accurate solution of the
sum rules (4.13). We are primarily interested in the energy region not
very far from the inelastic threshold, i.e.,
$$
(M_h + \e_0)\ \lsim \l\ \lsim\ 5\ {\rm GeV}, \eqno (4.17)
$$
since we want to calculate in particular the absorption of \F's
in confined hadronic matter. In such an environment,
the constituents will be hadrons with momenta of at most
a GeV or two. A usual hadron ($\pi,~\rho$, nucleon) of 5 GeV momentum,
incident on a \J~at rest, leads to $\sqrt s \simeq 6$ GeV, and this
corresponds to $\l \simeq 5$ GeV.
\par
{}From what we learned above, the energy region corresponding to the range
(4.17) will be determined by the gluon structure function at
values of $x$ not far from unity. There
the $x$-dependence of $g(x)$ can be well described by a power law
$$
g(x) = g_2\ (k+1)\ (1-x)^k, \eqno(4.18)
$$
where the function (18) is normalized so that the second moment (4.12)
gives the fraction $g_2$ of the light hadron momentum carried
by gluons, $ <O_2> = g_2 \simeq 0.5$.
This suggests a solution of the type
$$
\sigma_{\f h}(y) = a (1-y)^{\alpha}, \eqno(4.19)
$$
where $a$ and $\alpha$ are constants to be determined.
Substituting (4.18) and (4.19) into the sum rule (4.13) and performing
the integrations, we find
$$
a\ {{\Gamma(\alpha+1)} \over {\Gamma(n+\alpha)}} =
\left({2 \pi^{3/2} g_2 \over m_Q^2}\right)\left(16 \over 3 \right)^2
\left( {16\pi \over {3g^2}}\right)
{\Gamma(n+{5 \over 2}) \over \Gamma(n+5)} {\Gamma(k+2) \over
\Gamma(k+n)}.
\eqno(4.20)
$$
We are interested in the region of low to moderate
energies; this corresponds to relatively large $x$, to which higher
moments are particularly sensitive.
Hence for the
range of $n$ for which Eq.\ (4.5) is valid \ref{Novikov},
$n~\lsim~8$, the essential
$n$-dependence is contained in the $\Gamma$-functions.
For $n~\gsim~4$, Eq.\ (4.20) can be
solved in closed form by using an appropriate approximation for the
$\Gamma$-functions. We thus obtain
$$
a\ {\Gamma(\alpha +1) \over \Gamma(k+2)} \simeq {\rm const.}\
n^{\alpha - k -5/2}. \eqno(4.21)
$$
Hence to satisfy the sum rules (4.14), we need
$$
\alpha=k+{5 \over 2}~~~~~~~
a = {\rm const.}\ {\Gamma(k+2) \over \Gamma(k+{7 \over 2})}. \eqno(4.22)
$$
Therefore the solution of the sum rules (4.13) for
moderate energies $\lambda$ takes the form
$$
\sigma_{\f h}(\lambda) = 2 \pi^{3/2} g_2  \left(16 \over 3 \right)^2
\left( {16\pi \over {3g^2}}\right){1 \over m_Q^2}
{\Gamma(k+2) \over \Gamma(k+{7 \over 2})}
\left(1-{\l_0 \over \l}\right)^{k+5/2}. \eqno(4.23)
$$
To be specific, we now consider the \J-nucleon interaction.
Setting $k=4$ in accord with quark counting rules, using
$g_2\simeq 0.5$ and expressing the strong coupling $g^2$ in terms of
the binding energy $\e_0$ (Eq.\ 4.10),
we then get from Eq.\ (4.23) the energy dependence of
the $\j~\!N$ total cross section
$$
\sigma_{\j N}(\l) \simeq 2.5\ {\rm mb} \times
\left(1-{\l_0 \over \l}\right)^{6.5}, \eqno(4.24)
$$
with $\l$ given by Eq.\ (6) and $\l_0 \simeq (M_N+\e_0)$.
This cross section rises very slowly from threshold, as shown in Fig.\
4.2; for $P_N \simeq 5 $ GeV, it is around 0.1 mb, i.e., more than an
order of magnitude below its asymptotic value.
\par
We should note that the high energy cross section of 2.5 mb in Eq.\
(4.24) is {\sl calculated} in the short-distance formalism of QCD and
determined numerically by the values of $m_c$ and $\e_0$.
{}From Eqs.\ (4.10) and (4.23), it is seen to be proportional to
$1/(m_Q\sqrt{m_Q \e_0})$. For $\u-N$ interactions, with $m_b\simeq$ 4.5
GeV and $\e_0 \simeq 1.10$ GeV, we thus have the same form (4.24),
but with
$$
\sigma_{\u N} \simeq 0.37~{\rm mb} \eqno(4.25)
$$
as high energy value. This is somewhat smaller than that obtained from
geometric arguments \ref{Povh} and potential theory \ref{MT}.
\par
For sufficiently heavy quarks, the dissociation of quarkonium states
by interaction with light hadrons can thus be fully accounted for
by short-distance QCD. Such perturbative calculations become valid
when the space and time scales associated with the quarkonium state,
$r_Q$ and $t_Q$, are small in comparison to the nonperturbative
scale $\la^{-1}$
$$
r_Q<<\la^{-1}, \eqno(4.26a)
$$
$$
t_Q<<\la^{-1}; \eqno(4.26b)
$$
$\la^{-1}$ is also the characteristic size of the light hadrons.
In the heavy quark limit, the quarkonium binding becomes Coulombic,
and the spatial size $r_Q \sim (\alpha_sm_Q)^{-1}$ thus is small.
The time scale is by the uncertainty relation given as the inverse
of the binding energy $E_Q \sim m_Q$ and hence also small.
For the charmonium ground state $\j$, we have
$$
r_{\j} \simeq 0.2~{\rm fm}= (1~ {\rm GeV})^{-1};~~~
E_{\j} = 2 M_{D} - M_{\p} \simeq 0.64~{\rm GeV}. \eqno(4.27)
$$
With $\la \simeq 0.2$ GeV, the inequalities (4.25) seem already
reasonably well satisfied, and also the heavy quark relation
$E_{\j} = (1/m_c r_{\j}^2)$ is very well fulfilled.
We therefore expect that the dissociation of \J's in hadronic matter
will be governed by the \J-hadron break-up cross
section as calculated in short-distance QCD.
\par
Nevertheless, in view of the finite charm quark mass, it makes sense
to ask if the formalism developed here correctly describes
\J~interactions with light hadrons. We will take up this question first
theoretically, in the next section, where we shall see if
non-perturbative interactions can lead to significant contributions to
\J~break-up. In Section 4.3, we then see how to verify
experimentally the short-distance QCD prediction for \J~break-up
by light hadrons.
\medskip\noindent
{\sl 4.2~~Non-Perturbative Quarkonium Dissociation}
\medskip
For an isolated \J-hadron system, non-perturbative interactions can be
pictured most simply as a quark rearrangment \ref{Larry}. Consider
putting a \J~``into" a stationary light hadron;
the quarks could then just rearrange their binding pattern
to give rise to transitions such as $\j + N \to \Lambda_c+{\bar D}$ or
$\j+\rho \to D + {\bar D}$ (Fig.\ 4.3). The probability for such a
transition can be written as
$$
P_{\rm rearr} \sim \int d^3 r~R(r)~|\phi_{\psi}(r)|^2, \eqno(4.28)
$$
where the spatial distribution of the $\c$ bound state is
given by the squared wave function $|\phi_{\psi}(r)|^2$. The function
$R(r)$ in eq.\ (4.28) describes the resolution capability of the colour
field inside the light hadron. Its wave length is of order $\la^{-1}$,
and so it cannot resolve the charge content of very much smaller bound
states; in other words, it does not ``see" the heavy quarks in a
bound state of
radius $r_Q<< \la^{-1}$ and hence cannot rearrange bonds.
The resolution $R(r)$ will approach unity for $r\la >>1$
and drop very rapidly with $r$ for $r\la < 1$, in the
functional form
$$
R(r) \simeq (r\la)^n, ~~r\la < 1,   \eqno(4.29)
$$
with $n=2$ \ref{Low} or 3 \ref{Peskin}.
As a result, the integrand of eq.\ (4.28)
will peak at some distance $r_0$, with $r_Q< r_0 < \la$.
Since the bound state radius of the
quarkonium ground state decreases with increasing heavy quark mass,
while $R(r)$ is $m_Q$-independent, $r_0 \to r_Q \to 0$ as $m_Q \to
\infty$. Hence $P_r$ vanishes in the limit $m_Q \to \infty$
because $R(r_0)$ does, indicating that
the light quarks can no longer resolve the small heavy quark
bound state.
\bigskip
\vskip 6truecm
\medskip \centerline{Fig.\ 4.3: Rearrangement transition
$\rho + \j \to D + {\bar D}$}
\bigskip
In a potential picture, the situation just described means that the
charm quarks inside the
\J~have to tunnel from $r=r_{\psi}$ out to a distance at which the
light quarks can resolve them, i.e., out to some $r\simeq c \la^{-1}$,
where $c$ is a constant of order unity (Fig.\ 4.4).
Such tunneling processes are therefore
truly non-perturbative: they cover a large space-time
region, of linear size $\Lambda_{QCD}^{-1}$, and do not involve any
hard interactions. Following \ref{Larry}, we shall here estimate
the contribution of non-perturbative tunnelling to the dissociation
of quarkonium states.
\bigskip
\vskip 7.5 truecm
\medskip
\centerline{Fig.\ 4.4: \J~dissociation by tunneling}
\vfill\eject
In general, the problem of quark tunneling cannot be solved in a
rigorous way, since it involves genuine non-perturbative QCD
dynamics. However, the large mass of the heavy quark allows a
very important simplification, the use of the quasiclassical
approximation. In this approximation, the rate of tunneling
$R_{\rm tun}$ can be written down in a particularly transparent way:
it is simply the product of the frequency $\omega_{\p}$ of the heavy
quark motion in the potential well and the tunnelling probability
$P_{\rm tun}$ when the quark hits the wall of the well,
$$
R_{\rm tun} =\omega_{\psi} P_{\rm tun} \eqno(4.30)
$$
The frequency $\omega_{\psi}$ is
determined by the gap to the first radial excitation,
$$
\omega_{\psi} \simeq (M_{\psi'}-M_{\psi}) \simeq \ E_{\psi}. \eqno(4.31)
$$
Consider now the potential seen by the $\c$ (Fig.\ 4.4).
For a particle of energy $E$,
the probability of tunneling through the potential barrier $V(r)$
is obtained from the squared wave function in the ``forbidden" region.
It can be expressed in terms of the action $W$ calculated along the
quasiclassical trajectory,
$$
P_{\rm tun} = e^{-2W}, \eqno(4.32)
$$
where
$$
W= \int_{r_1}^{r_2} |p|\ dr, \eqno(4.33)
$$
Here the momentum $|p|$ is given by
$$
|p|=\left[ 2M(V(x)-E) \right]^{1/2}, \eqno(4.34)
$$
and $r_1, r_2$ are the turning points of the classical motion
determined from the condition $ V(r_i)=E $.
\par
In our case, the width of the barrier is approximately $0.6~\la^{-1}$,
while its height $(V - E)$ is equal to the dissociation threshold
$E_Q$. The mass $M$ in Eq.(4.34) is the reduced mass, $M=m_Q/2$.
We thus have
$$
W \simeq 0.6~\sqrt{m_Q\ E_Q}/\la. \eqno(4.35)
$$
For the \J, we get from (4.35) the value $W \simeq 3$; this
$a\ posteriori$ justifies the use of quasiclassical approximation,
which requires $S>1$.
\par
Using eq.\ (4.35), we obtain as final form for the tunneling rate
(4.30)
$$
R_{\rm tun} = E_{\psi} \
{\rm exp} - (1.2 \sqrt{m_c\ E_{\psi}}/ \la).
\eqno(4.36)
$$
With the above mentioned \J~parameters, this leads to the
very small dissociation rate
$$
R_{\rm tun} \simeq 9.0 \times 10^{-3}~{\rm fm}^{-1}.\eqno(4.37)
$$
In terms of $R_0$, the \J~survival probability is given by
$$
S_{\rm tun} = {\rm exp}\left\{
-\int_0^{t_{\rm max}} dt~R_{\rm tun}\right\}, \eqno(4.38)
$$
where $t_{\rm max}$ denotes the
maximum time the \J~spends adjacent to the light hadron. In the
limit $t_{\rm max} \to \infty$, $S_{\p}$ vanishes. However,
the uncertainty relations prevent a localisation of the two systems
in the same spatial area for long times. From $\Delta x \leq
\la^{-1}$ we get $\Delta p \geq \la$, so
that the longest time which the \J~can spend in the interaction
range of the light hadron is
$$
t_{\rm max} = \la^{-1} \left( 1 + {m^2 \over
\la^2} \right)^{1/2}, \eqno(4.39)
$$
with $m$ for the mass of the light hadron. For nucleons or vector
mesons, this time is 4 - 5 fm, and with this, the survival probability
is very close to unity; hence non-perturbative tunneling interactions
provide only negligible contributions to \J~dissociation.
\par
In addition to such tunnelling, there can be direct and sequential
thermal excitation to the continuum. In particular the latter still
requires further analysis \ref{Larry}.
\medskip\noindent
{\sl 4.3~~The Experimental Study of Charmonium-Hadron Interactions}
\medskip
We now return to the break-up rate for \J~interactions with slow
light hadrons, as predicted by heavy quark QCD. Since this prediction is
crucial for the use of quarkonia as confinement/deconfinement probe, it
certainly must be checked experimentally. We shall now consider
how that can be done.
\par
In Eq.\ (3.22), we
had obtained the survival probability for charmonium states $i$ in $p-A$
interactions in terms of the break-up cross sections $\sigma_i$. We now
just have to insert the cross section (4.24) into this expression
to obtain the \J~survival probability as function of $x_F$. In Fig.\ 4.5
the result is shown for a cms collision energy of 17.4 GeV, the value
provided in CERN SPS $Pb$-beam experiments. Included in this figure
is also the colour octet interaction region $x_F\geq 0$, together with
data \ref{Badier} and the fit obtained in section 3.2. We see that in
the regime $x_F\leq -0.2$, in which physical \J's interact with the
nuclear medium, the survival probability is essentially unity, due to
the smallness of the cross section (4.24) in the threshold region.
\par
This picture has to be contrasted to the geometric absorption approach,
which provides the basis for all hadronic accounts of \J-suppression
in nucleus-nucleus collisions. Here the \J~cross section is assumed to
attain its high energy value $\sigma_{\j N}(s=\infty) \simeq 2.5$ mb
immediately at threshold. The survival probability for this case is also
shown in Fig\ 4.5, and is seen to differ both qualitatively and
quantitatively from that based on the QCD result (4.24). Through a
measurement of \J-production in $p-A$
collisions at negative $x_F$, technically attainable through data from a
$Pb$-beam incident on a hydrogen or deuterium target at positive
$x_F$, one can thus check if the threshold behaviour predicted by
short distance QCD for inelastic \J-hadron interactions is indeed
correct.
\par
The actual experimental results will differ from what is shown in
Fig.\ 4.5 for two reasons. We have already noted in section 3.4 the
modifications expected because of the EMC effect. A second reason is
that the \J~peak observed in the measured dilepton spectrum is only to
about 60 \% due to directly produced $1S$ $\c$ \J~resonances; the
remaining 40 \% are mainly due to $\chi_c$ production with the
subsequent radiative decay $\chi_c \to \j + \gamma$
\refs{Lemoigne}{Anton1}.\footnote{*}{We
shall for simplicity ignore here a further
small contribution ($\lsim~5$ \%) fom \P~decay).}
\par
Concerning the EMC effect, we can either fold it into the predictions
shown in Fig.\ 4.5, or it can be measured independently and then removed
from the \J-production data. Measuring Drell-Yan
dilepton production in the same kinematic region provides
$R_{A/p}(x)$ (see Eq.\ (3.23)) for quarks
directly, without any final state modification. A measurement of open
charm production leads to $R_{A/p}(x)$ in just the same
superposition of quark-antiquark annihilation and gluon fusion as in
charmonium production, but again without any final state effect.
Separate measurements of Drell-Yan and/or open charm production would
thus determine the EMC modification without addititional final state
effects. Such measurements can therefore be used to unfold the EMC
modification of \J~data.
\par
To avoid any $\chi_c$ admixture, it would of course be ideal to measure
both \J~and $\chi_c$-production directly in $p-A$ collisions
\ref{Anton1}.
This may be too difficult due to the abundance of photons in the case
of heavy nuclei as targets. We shall therefore try to estimate the
effect of the $\chi_c$ admixture in the two scenarios considered
here by simply adding the corresponding predictions with the noted
60/40 weights. For the geometric approach with its asymptotic cross
sections, this requires a calculation completely analogous to that for
the \J, but now with $\sigma_{\chi} \simeq 6$ mb and the
correspondingly longer resonance formation time in the ${\bar L}$ of Eq.
(3.22).
\par
The application of short distance QCD to calculate the inelastic
$\chi_c$-hadron cross section is as reliable as for the \J,
since the binding energy of the $\chi_c$, $\e_{\chi}
\simeq$ 0.24 GeV, is about equal to $\la$. Nevertheless, such a
calculation can give us some idea of the expected behaviour. The result
is \ref{KS5}
$$
\sigma_{\chi N}({\bar s}) \simeq 11.3~{\rm mb} \times \left( 1 -
\left[{
2M_{\chi}(m+\epsilon_{\chi}) \over ({\bar s}- M_{\chi}^2)}\right]
\right)^{6.5}.
\eqno (4.40)
$$
The asymptotic value is thus a factor two
larger than the geometric estimate; this is a consequence of the fact
that short distance QCD \refs{Bhanot}{SVZ} leads to higher
powers in the bound state radii than just $r^2$. The behaviour of the
$\chi_c  N$ cross section (4.40) is shown in Fig.\ 4.2. Inserting
Eq.\ (4.40) into (3.22) and adding \J~and $\chi_c$ contributions then
gives
us the short distance QCD survival probability for the measured \J. It
and the corresponding result from the geometric approach are shown in
Fig.\ 4.6. As seen, for $x_F\leq -0.4$, the two approaches differ
qualitatively in their functional form and quantitatively by more than
20\%. An experimental test of the theory for the interaction of heavy
quarkonia with light hadrons should therefore be possible.
\par
As last point in this section, we comment briefly on the interaction of
the \P~with light hadrons. Since its binding energy is only about 50
MeV, it lies almost at the open charm threshold and can definitely
not be treated by short distance QCD. Here it might therefore not be
unreasonable to assume that it attains its high energy value rather soon
after threshold. This value can only be estimated by geometric
arguments, and these suggest around 10 mb \refs{Povh}{MT}.
\bigskip\noindent
{\bf 5.\ Quarkonium Dissociation in Confined and Deconfined Media}
\medskip
Here we first show that at fixed temperature (or energy density), a
deconfined medium contains much harder gluons than a confined medium.
Tightly bound quarkonia probe gluon hardness: while they were
found to remain essentially unaffected in a confined medium at
$T~\lsim~0.5$ GeV, a QGP of such temperature is shown to be very
effective in their disssociation. To relate this to the environment
produced in nuclear collisions, the resulting charmonium survival is
studied in the case of isentropic longitudinal expansion.
\medskip\noindent
{\sl 5.1~~The Parton Structure of Confined vs. Deconfined Matter}
\medskip
The ultimate constituents of matter are evidently always quarks and
gluons. What we want to know is if these quarks and gluons are
confined to hadrons or not. Let us therefore assume that we are given a
macroscopic volume of static strongly interacting matter and have to
analyse its confinement status.
\par
As prototype for matter in a confined state, we consider an ideal gas
of pions. Their momentum distribution
is thermal, i.e., for temperatures not too low it is given by
${\rm exp}(-E_{\pi}/T) \simeq {\rm exp}(-p_{\pi}/T)$. Hence the average
momentum of a pion in this medium is $\langle p_{\pi} \rangle = 3~T$. The
distribution of quarks and gluons within a pion is known from structure
function studies; the gluon density is $g(x) \simeq 0.5 (1-x)^3$ for
large $x=p_g/p_{\pi}$.\footnote{$^{1)}$}{For very small $x$, recent
results
from HERA indicate a steeper increase; however, this does not affect our
considerations here.} As a consequence, the average momentum of a gluon
in confined matter is given by
$$
\langle p_g \rangle_{\rm conf} = {1 \over 5} \langle p_{\pi} \rangle =
{3 \over 5} T. \eqno(5.1)
$$
Hence in a medium of temperature $T \simeq 0.2$ GeV, the average gluon
momentum is around 0.1 GeV.
In contrast, the distribution of gluons in a deconfined medium is
directly thermal, i.e., ${\rm exp}(-p_g/T)$, so that
$$
\langle p_g \rangle_{\rm deconf} = 3T. \eqno(5.2)
$$
Hence the average momentum of a gluon in a deconfined medium is five
times higher than in a confined medium\footnote{$^{2)}$}{We could
equally well
assume matter at a fixed energy density, instead of temperature. This
would lead to gluons which in case of deconfinement are approximately
three times harder than for confinement.}; for $T=0.2$ GeV, it
becomes 0.6 GeV. An immediate consequence of deconfinement is thus
a considerable hardening of the gluon momentum distribution. Although we
have here presented the argument for massless pions as hadrons, it
remains essentially unchanged for heavier mesons ($\rho/\omega$) or
nucleons, where one can use a non-relativistic thermal distribution for
temperatures up to about 0.5 GeV.
We thus have to find a way to detect such a hardening of the gluon
distribution in deconfined matter.
\par
The lowest charmonium state \J~provides an ideal probe for
this. It is very small, with a radius $r_{\psi}\simeq 0.2$ fm $\ll
\Lambda^{-1}_{QCD}$, so that \J~interactions with the conventional light
quark hadrons probe the short distance features, the parton
infra-structure, of the latter. It is very strongly bound, with a
binding energy $\epsilon_{\psi} \simeq 0.65$ GeV $\gg \Lambda_{QCD}$;
hence it
can be broken up only by hard partons. Since it shares no quarks or
antiquarks with pions or nucleons, the dominant perturbative interaction
for such a break-up is the exchange of a hard gluon, and this was the
basis of the short distance QCD calculations presented in the previous
chapter.
\par
We thus see qualitatively how a deconfinement test can be carried out.
If we put a \J~into matter of temperature $T=0.2$ GeV, then
\item{--}{if the matter is confined, $\langle p_g \rangle_{\rm conf}
\simeq 0.1$ GeV, which is too soft to resolve the \J~as a
$\c$ bound state and much less than the binding energy
$\epsilon_{\psi}$, so that the \J~survives;}
\item{--}{if the matter is deconfined, $\langle p_g \rangle_{\rm decon}
\simeq$ 0.6 GeV, which (with some spread in the momentum distribution)
is hard enough to resolve the \J~and to break the binding, so that the
\J~will disappear.}
\par\noindent
The latter part of our result is in accord with the mentioned prediction
that the formation of a QGP should lead to \J~suppression
\refs{Matsui}{Hwa}. There it was argued that in a QGP, colour screening
would prevent any
resonance binding between the perturbatively produced $c$ and ${\bar
c}$, allowing the heavy quarks to separate. At the hadronisation point
of the medium, they would then be too far apart to bind to a \J~and
would therefore form a $D$ and a $\bar D$. Although the details of such
a picture agreed well with the observed \J~suppression \ref{GS2}, it
seemed possible to obtain a similar suppression by absorption in a
purely hadronic medium \ref{Blaizot/Hwa}, through collisions of the
type
$$
\j + h \to D + {\bar D} + X. \eqno(5.3)
$$
Taking into account the partonic substructure of such hadronic
break-up processes, we now see that this is in fact not possible for
hadrons of reasonable thermal momentum. Our picture thus not only
provides a dynamical basis for \J~suppression
by colour screening, but it also indicates that in fact \J~suppression
in dense matter will occur {\sl if and only if} there is deconfinement.
We note, however, that the dynamical approach to \J~suppression does not
require a thermal equilibration of the interacting gluons, so that it
will remain applicable even in deconfined pre-equilibrium stages.
\par
While we have studied the hadronic part of the argument in detail in the
previous sections, we have so far not considered the dynamics of
quarkonium dissociation in a deconfined medium. This will be taken up in
the next section.
\medskip\noindent
{\sl 5.2~~Quarkonium Dissociation by Deconfined Gluons}
\medskip
In section 4.2 we had obtained the cross section for the dissociation of
a tightly bound quarkonium by an incident light hadron. Eq.\ (4.23) was
obtained \ref{KS3} by convolution of the inelastic gluon-charmonium
cross section with the gluon distribution in the light hadron. The
gluon-quarkonium cross section itself is given by
$$
\sigma_{g\Phi}(k) = {2\pi \over 3} \left( {32 \over 3} \right)^2 \left(
{m_Q \over \e_0} \right)^{1/2} {1 \over m^2_Q} {(k/\e_0 - 1)^{3/2}
\over (k/\e_0)^5 }, \eqno(5.4)
$$
with $k$ denoting the momentum of the gluon incident on a stationary
quarkonium. In Fig.\ 5.1 we show the resulting break up cross section
for gluon-\J~and gluon-\U~interactions as function of the gluon
momentum $k$. They are seen to be broadly peaked in the range $0.7 \leq
k \leq 1.7$ GeV for the \J, with a maximum value of about 3 mb, and in
the range $1.2 \leq k \leq 2.2$ GeV for the \U, with a maximum of about
0.45 mb. Also shown in Fig.\ 5.1 are the corresponding cross sections
for incident pions (note that now $k=3$ in Eq.\ (4.2)), with high energy
values of 3 mb and 0.5 mb for \J~and \U, respectively. The break-up by
incident hadrons is seen to be negligible up to momenta of around 4 GeV
for
the \J~and 7 GeV for the \U. Fig.\ 5.1 thus provides the basis for the
claim that in matter temperature $T\leq 0.5$ GeV, gluons of thermal
momentum can break up charmonia, while hadrons cannot.
We note here that, just as in the photoelectric dissociation of atoms,
the break-up is most effective when the momentum of the gluon is
somewhat above the binding energy. Gluons of lower momenta can neither
resolve the constituents in the bound state nor raise them up to the
continuum; on the other hand, those of much higher
momenta do not see the (by their scales) large object and just pass
through it.
\par
To illustrate this more explicitly, we calculate the break-up cross
section for the \J~as function of the temperature $T$ of an ideal QGP.
Using Eq.\ (5.4) with $m_c=1.5$ GeV and the \J~binding energy of 0.64
GeV, we then get
$$
\sigma_{g\j}(T) = 65~{\rm mb} \times { \int_{\e_0}^{\infty} dk~k^2
e^{-k/T}
(k/\e_o-1)^{3/2}
(k/\e_0)^5 \over \int_{\e_0}^{\infty} dk~k^2 e^{-k/T}}. \eqno(5.5)
$$
The result is shown in Fig.\ 5.2 and confirms that up to about $T\sim
0.5$ GeV, only a deconfined medium can dissociate \J's. We see moreover
that the effective cross section for break-up in the temperature range
$0.2 \leq T \leq 0.5$ GeV is about 1.2 mb. It is this value which will
determine the suppression of the (pure $1S$) \J~in a deconfined medium.
\par
Before we turn to charmonium production in a more realistic non-static
environment, we want to consider briefly the possible role of quarks in
the dissociation of charmonia in a QGP. This can be treated in a fashion
quite similar to the interaction of quarkonia with light hadrons; the
gluon distribution function in the hadron should now be replaced by an
effective gluon distribution ``in" quarks, i.e., the quark splitting
function $P(x)$ characterising the process $ q \to q + g$. It's
functional form for $x\to 1$
is fixed both by quark counting rules \ref{counting}
and the Altarelli-Parisi equations \ref{splitting},
$$
P(x) \sim (1-x)^2, \eqno(5.6)
$$
which leads to an average gluon momentum
$$
\langle p_g \rangle_{\rm deconf}^q = {3 \over 4} T. \eqno(5.7)
$$
The average momentum of gluons emitted by thermal quarks is thus
higher than for gluons confined to thermal
hadrons; it is nevertheless low enough to suggest that in an
equilibrium plasma, the direct interaction with thermal
gluons is the main dissociation mechanism. In pre-equilibrium,
however, the quarks can be much faster and become the
dominant cause of dissociation.
\medskip\noindent
{\sl 5.3~~Charmonium Survival in an Expanding Medium}
\medskip
The exponential quarkonium survival probability of the general form
(3.20) applies to a stationary medium of finite size. In nuclear
collisions, the medium needs some proper time $t_0$ to be formed, and it
then expands until a time $t_f$ when the era of strong interactions
in the medium
ends. Hence slow quarkonia in the medium of not too high an initial
temperature will in general stop interacting when the medium has cooled
down enough, even though they have not yet left it.
The survival probability of a quarkonium state $i$ in such an
expanding medium can be written as
$$
S(T_0) = exp\{ - \int_{t_0}^{t_f} dt~n(t)~\sigma_i(t)\}, \eqno(5.8)
$$
where $n(t)$ is the density of scattering centers at time $t_0 \leq t
\leq t_f$ and $\sigma_i(t)$ the break-up cross section for state $i$ in
the medium at that time. If we assume isentropic longitudinal expansion,
density, temperature and time are related by
$$
{n(t_o) \over n(t)} = \left( {T_0 \over T} \right)^3 = {t \over t_0},
\eqno(5.9)
$$
with $T \leq T_0$. Using this relation, we can rewrite Eq.\ (5.6) in the
form
$$
S(T_0) = exp\{ -3 n_h(T_0) t_0 \int_{T_f}^{T_c} dT {\sigma_i^h(T) \over
T} - 3 n_g(T_0) t_0 \int_{T_c}^{T_0} dT {\sigma_i^g(T) \over T} \},
\eqno(5.10)
$$
where $T_f$ is the temperature of the medium for which the break-up of
state $i$ stops. The first term in the exponent corresponds to
dissociation by hadrons, the second by gluons. We had seen that the main
contribution to the dissociation comes from gluon-quarkonium
interactions, making $S(T_0) \simeq 1$ up to $T_0\simeq T_c \simeq
0.15$ GeV. Hence
by inserting the cross section (5.5) (Fig.\ (5.2)) into Eq.\ (5.10), we
obtain the survival probability of a \J~in a medium in isentropic
longitudinal expansion, for an initial temperature $T_0 \geq T_c$;
below that, it is unity. The thermalisation time
$t_0$ is generally argued to be around 1 fm; and for an ideal QGP, the
density of gluons at temperature $T$ is given by the Stefan-Boltzmann
form
$$
n(T) = {8 \pi^2 \over 45} T^3. \eqno(5.11)
$$
The resulting suppression as function of temperature is shown in
Fig.\ 5.3. The survival probability is seen to decrease very rapidly
with increasing temperature, essentially vanishing for $T~\gsim~0.3$
GeV.
\bigskip\noindent
{\bf 6.\ Pre-Equilibrium Deconfinement}
\medskip
In the previous chapter, we established quarkonium suppression as a
probe to check if a given sample of strongly interacting {\sl matter}
consists of deconfined quarks and gluons. In an actual nuclear collision,
however, such a suppression could have been caused before the onset of
thermalisation. Here we therefore study the possibility of
pre-equilibrium quarkonium suppression.
\medskip\noindent
{\sl 6.1~~Shadowing and \J~Suppression in Nucleus-Nucleus Collisions}
\medskip
Consider a nucleus-nucleus collision at CERN SPS energy ($\sqrt s \simeq
20$ GeV in the nucleon-nucleon cms), leading to \J~production at
mid-rapidity, $y_{\j}$=0, with the production mechanism as described
in Chapter 2. During
the first 0.25 fm, the produced $\c$ pair is a colour octet and will
interact as such with the passing nucleons of target and projectile.
The momenta of these nucleons in the rest system of the \J~will be
around 10 GeV. The situtation seen by the \J~is thus very similar
to that encountered at $x_F=0$ in collisions of 200 GeV/c protons
with a nuclear target. As discussed in Chapter 3, one here finds
a suppression dominated by ``nuclear shadowing", i.e., destructive
interference of the scattering on different nucleons. Such an effect
will now arise from both projectile and target, however, and this
suppression must be removed before final state effects can be studied
\ref{GS1}.
\par
To see the effect of this shadowing correction, we show in Fig.\ 6.1
the latest data taken by the NA38 collaboration at CERN in $p-U$ and
$S-U$ collisions \refs{Ronceux}{Carlos}. Here the suppression is
measured with respect to the high mass Drell-Yan continuum, isospin
corrected for $p-U$. The nucleus-nucleus
data are shown as function of the neutral transverse
hadronic energy $E_T^0$ produced in the collision. The
$p-U$ value is simply included in this figure; it is
not associated to any particular $E_T^0$. The difference between
the $p-p$ and $p-A$ results is, as discussed in Chapter 3, essentially
given by the shadowing function $R_{A/p}^{sh}(x_F\simeq 0)$.
To correct the data for shadowing effects, we therefore divide the
$p-U$ data by $R_{U/p}^{sh}(0)$ and the $S-U$ data by
$R_{S/p}^{sh}(0)\times R_{U/p}^{sh}(0)$.
The result is shown in Fig.\ 6.2; the $p-U$ value and that for the
$S$ beam at low $E_T^0$ now agree.
The remaining suppression in the nuclear collision data is now due
to effects on the colour singlet $\c$ in its state after the time of
colour neutralisation.
\par
We thus want to study the effect of the environment on this
$\c$. If we ignore nuclear stopping, target and
projectile nucleons retain in the $\c$ cms their initial
momenta $P_N \simeq 10$ GeV/c. The $\c$ then sees two nuclei, each
Lorentz-contracted to about 1 fm, since $(2m/\sqrt s) \simeq 0.1$; for
simplicity, we consider here $A-A$ collisions.
At the colour neutralisation time $\t_8 \simeq 0.25$ fm, the centers of
these thin discs have become separated by approximately 0.5 fm, so that
the target and projectile nucleons still have considerable overlap.
Stopping would slow down the nucleons to increase this. The detailed
kinematics encountered by the colour singlet $\c$ is, however, quite
unimportant for the essential question: can a pre-equilibrium state of
hadrons account for the observed suppression? The initial medium, at the
point of maximum overlap between target and projectile, is one of twice
nuclear density, containing nucleons of 10 GeV/c momentum in the $\c$
rest system; equivalently, we can assume normal nuclear density and an
effective $\c$ path length $L = 2\times 3R_A/4$,
with $R_A=1.12~A^{1/3}$ \ref{Gerschel}.
Whatever scattering processes now occur lead to an increase
of the density $n$, but at the same time to a decrease of the momenta
$P$ of the hadronic constituents of the medium. Momentum conservation
requires for the momentum flow through a given surface $F$
$$
P~n~F = P_N n_0 F = {\rm const.} \eqno(6.1)
$$
so that the momentum in fact drops as $1/n$. If we neglect the cross
section reduction during the time
needed for the singlet $\c$ to become a full-sized resonance, the
survival probability of a \J~in the medium becomes
$$
S_{\j} = exp\{- n_0 \sigma(P_N) L\}. \eqno(6.2)
$$
The medium actually seen by the \J~will have a higher density
\ref{Gavin}, but its constituents will have lower momenta. In view of
Eq.\ (6.1), we can nevertheless use the form (6.2) to calculate the
survival probability. In contrast to the phenomenological fit of
\ref{Gerschel}, Eq.\ (6.2) contains the break-up cross section
at the actual collision energy, and in
Chapter 4 we have seen that this cross section
is much below its geometric high energy
value. At $P_N = 10$ GeV/c, we find (see Fig.\ 4.2) $\sigma_{\j}\simeq
0.8$ mb. Inserting this into Eq.\ (6.2), with an average value
$L \simeq 10$ fm for central $S-U$ collisions, leads to
$S\simeq 0.9$ for the survival probability. Since we have here
neglected the cross section reduction during the \J~evolution
as well as the path length reduction by the part already
included in the shadowing correction, the actual
survival probability will be larger. Thus hadronic pre-equilibrium
interactions cannot account for the measured \J~suppression.
\par
We therefore conclude that the \J~suppression observed in the NA38
experiment \ref{NA38} provides evidence for the existence of {\sl
deconfined} gluons in the medium probed by the \J. By this we mean
that the medium
in which the \J~finds itself after a central nucleus-nucleus collision
contain gluons whose momentum distribution is harder than that found in
mesons or nucleons \ref{KS6}. Only in collisions with such gluons
can the tightly bound \J~be dissociated; gluons confined to hadrons
are not sufficiently hard. We emphasize that this conclusion makes
crucial use of the energy dependence of the \J-light hadron break-up
cross section as calculated in short distance QCD. As pointed out in
Chapter 4, this result can and should be experimentally checked.
\medskip\noindent
{\sl 6.2~~Deconfined Gluons and QGP}
\medskip
The existence of deconfined gluons is clearly {\sl not} equivalent to
the existence of a QGP, in which such gluons are in thermal
equilibrium. It is only the first step towards a QGP: it shows that in
the large-scale medium there exist deconfined partons. Their
thermalisation comes at the end of a parton interaction cascade, and
it is not at all clear whether at present energies there are
enough deconfined partons in a sufficiently large and long-lived
system to reach this stage. How can we check experimentally whether
parton thermalisation is or is not achieved?
\par
The cascade formation of a QGP \refs{Mueller}{Wang} starts with the
production of gluons in primary collisions; these then interact again to
produce secondary gluons, quite possibly through multigluon production
\ref{Shuryak}, and so on, until production and absorption
balance to form an equilibrium system. In equilibrium, the number of
gluons per unit volume is simply determined by the energy (or entropy)
density of the system. In the pre-equilibrium stage, it is lower and
(still) proportional to the number of primary collisions. We thus expect
\J~suppression by deconfined gluons in a pre-equilibrium system to
increase with the number of nucleon-nucleon collisions (or
equivalently, with the effective path length of the \J). On the other
hand, equilibrium suppression would be independent of the number of
such collisions and depend on the effective energy density of the system
only \ref{KS6}.
\par
To test this, we can study $Pb-Pb$ collisions as function of increasing
centrality (i.e., of increasing $E_T$). In this case, the energy
density remains essentially unchanged \ref{Karsch} while the number of
collisions increases.\footnote{*}{Since a change in centrality of $S-U$
collisions changes both the number of primary collisions and the
effective energy density, this does not distinguish pre-equilibrium from
equilibrium.} If the \J~suppression is found to {\sl increase},
the system cannot have reached equilibrium, and hence the suppression
must be due to deconfined gluons in the pre-equilibrium stage. Once the
suppression becomes independent of the number of primary collisions,
equilibrium and hence the QGP is reached. The forthcoming NA50 $Pb$-beam
data from the CERN-SPS should thus be able to resolve this question.
\bigskip\noindent
{\bf 7. Conclusions}
\medskip
We first summarize the essential theoretical points of this work, and
then survey the main conclusions we think can be drawn from present
data.
\par
Quarkonium production in hadron-hadron collisions is today quite well
understood in terms of elementary parton interactions (gluon fusion,
quark-antiquark annihilation). The distributions of the partons within
the colliding hadrons are determined in deep inelastic lepton-hadron
scattering, and the (non-perturbative) colour neutralisation of the
produced heavy \Q~pair can also be fixed empirically.
\par
Hadron-nucleus collisions determine what happens to quarkonium
production in a confined medium. In the production of fast quarkonia in
the nuclear rest frame, a fast colour octet \Q~passes the medium,
leading to quantum-mechanical interference (``nuclear
shadowing") and energy loss. Slow quarkonia in the nuclear rest frame
are subject to EMC effect modifications of the colliding partons in
addition to collisions with nucleons in the nucleus.
\par
Because of the small size and the large binding energy of the lowest
quarkonium states, their interaction with light hadrons is calculable in
short distance QCD. They can interact in leading
order only through the exchange of a hard gluon, and the gluon
distribution in the light hadrons is known to be very soft. The resulting
prediction is a cross section which rises very slowly from threshold
to its high energy value, suppressing strongly any break-up of
quarkonium ground states by slow mesons or nucleons.
\par
As a consequence of this suppressed quarkonium dissociation by light
hadrons, confined matter (at temperatures $T\simeq 0.5$ GeV) becomes
transparent to \J's. The momentum of deconfined thermal gluons, on the
other hand, is large enough to give rise to effective \J~dissociation;
such dissociation can occur also by deconfined gluons not in
equilibrium.
Strongly interacting matter thus leads to \J~suppression if and only if
it is deconfined. The loosely bound \P~can be broken up in both confined
and deconfined matter, though presumably more in a deconfined medium.
\par
In nucleus-nucleus collisions, the interaction of the coloured $\c$ pair
with target and projectile nucleons can be taken into account through the
nuclear shadowing determined in hadron-nucleus collisions. Any strong
\J~suppression remaining after removal of these shadowing effects is
accountable only by interaction with deconfined gluons.
\par
What have we then learned from the $h-A$ \refs{Badier}{E772} and $A-B$
\refs{NA38}{NA38-new} data available so far? What should the forthcoming
$Pb$-beam data clarify?
\par
The equality of \J~and \P~suppression in $h-A$ collisions for $x_F\geq
0$, as well as the size and $x_F$ dependence of the observed effect, are
in full
accord with the passage of a colour octet through nuclear matter. The
equality of \J~and \P~suppression and the observed $x_F$ dependence
are in clear disagreement with any description based on
the absorption of physical charmonium states in nuclear matter.
\par
There is a lack of data for charmonium production in a kinematic regime
in which fully formed \J's could interact with nuclear matter. Such data
could be obtained by experiments using the $Pb$-beam incident on a light
target \ref{KS5}.
\par
The observed difference between \J~and \P~suppression in nucleus-nucleus
interactions \ref{NA38-new} indicates that the relevant mechanism here
is different from that in $h-A$ collisions.
\par
The observed \J~suppression cannot be accounted for in terms of
nuclear shadowing on both projectile and target; even after removal of
shadowing, an $E_T$-dependent suppression (about 40 \% between low and
high $E_T$) remains.
\par
If the strong threshold suppression of \J~break-up by light hadrons, as
predicted in short distance QCD, is experimentally confirmed, the
\J~suppression observed in $O-U$ and $S-U$ interactions can only be
accounted for by the presence of deconfined gluons.
\par
It remains open whether the deconfined gluons required for
\J~suppression in nuclear collisions are already equilibrated;
hence their existence does not establish QGP formation, but quite
likely only a first step towards a thermalised deconfined medium. If the
gluons are not in equilibrium, the resulting \J~suppression should
increase with increasing $E_T$ in $Pb-Pb$ collisions at fixed energy;
in equilibrium,
the suppression would remain approximately constant.
\bigskip\noindent
{\bf Acknowledgements}
\medskip
We thank L. McLerran, G. Schuler and R. Vogt for helpful and stimulating
discussions. The financial support of the German Research Ministry
(BMFT) under contract 06 BI 721 is gratefully acknowledged.
\vfill\eject
\noindent
{\bf References}
\medskip
\parindent=13pt
\item{\reftag{Matsui})}{T. Matsui and H. Satz, \PL B 178 (1986)
416.}
\par
\item{\reftag{KS3})}{D. Kharzeev and H. Satz, \PL B 334 (1994) 155.}
\par
\item{\reftag{Bj})}{J. D. Bjorken, ``Energy Loss of Energetic
Partons in Quark-Gluon Plasma", Fermilab-Pub-82/59-THY, August 1982
(unpublished), and Erratum.}
\par
\item{\reftag{Miklos})}{M. Gyulassy and M. Pl\"umer, \PL B 243
(1990) 432.}
\par
\item{\reftag{Jets})}{K. Eskola and X.-N. Wang, ``High $p_T$ Jet
Production in $p-p$ Collisions", in {\sl
Hard Processes in Hadronic Interactions}, H. Satz and X.-N. Wang (Eds.)}
\par
\item{\reftag{Quarko})}{R. V. Gavai et al., ``Quarkonium
Production in Hadronic Collisions", in {\sl
Hard Processes in Hadronic Interactions}, H. Satz and X.-N. Wang
(Eds.); CERN Preprint CERN-TH.7526/94 (December 1994).}
\par
\item{\reftag{Miklos/XNW})}{M. Gyulassy et al., \NP A 538 (1992) 37c.}
\par
\item{\reftag{KS1})}{D. Kharzeev and H. Satz, \ZP C 60 (1993) 389.}
\par
\item{\reftag{Gyu-Wa})}{M. Gyulassy and X.-N. Wang, \NP B 420 (1994)
583.}
\par
\item{\reftag{Baier})}{R. Baier, Yu. Dokshitser, S. Peigne and D. Schiff,
``Induced Gluon Radiation in
a Deconfined Medium", Bielefeld Preprint BI-TP-94-57, November 1994.}
\par
\item{\reftag{GS1})}{S. Gupta and H. Satz, \ZP C 55 (1992) 391.}
\par
\item{\reftag{KS2})}{D. Kharzeev and H. Satz, \PL B 327 (1994) 361.}
\par
\item{\reftag{Cley})}{J. Cleymans et al., ``Prompt Photon
Production in $p-p$ Collisions", in {\sl
Hard Processes in Hadronic Interactions}, H. Satz and X.-N. Wang (Eds.)}
\par
\item{\reftag{VesaHP})}{S. Gavin et al., ``Production of Drell-Yan
Pairs in High Energy Nucleon-Nucleon Collisions", in {\sl
Hard Processes in Hadronic Interactions}, H. Satz and X.-N. Wang (Eds.)}
\par
\item{\reftag{NA38})}{C. Baglin et al., \PL B220 (1989) 471; B251
(1990) 465, 472; B225 (1991) 459.}
\par
\item{\reftag{Hwa})}{H. Satz in {\sl Quark-Gluon Plasma},
R. C. Hwa (Ed.), World Scientific, Singapore 1990.
\par
\item{\reftag{CE})}{M. B. Einhorn and S. D. Ellis, \PR D12 (1975)
2007;\hfill\par
H. Fritzsch, \PL 67B (1977) 217;\hfill\par
M. Gl\"uck, J. F. Owens and E. Reya, \PR D17 (1978) 2324; \hfill\par
J. Babcock, D. Sivers and S. Wolfram, \PR D18 (1978) 162.}
\par
\item{\reftag{MRS})}{A. D. Martin, R. G. Roberts and W. J. Stirling, \PL
B 306 (1993) 145.}
\par
\item{\reftag{GRV})}{M. Gl\"uck, E. Reya and A. Vogt, \ZP C53 (1993)
127.}
\par
\item{\reftag{Craigie})}{N. Craigie, Phys. Rep. 47 (1978) 1.}
\par
\item{\reftag{B-R})}{R. Baier and R. R\"uckl, \ZP C 19 (1983) 251.}
\par
\item{\reftag{Lemoigne})}{Y. Lemoigne et al., \PL B 113 (1982) 509.}
\par
\item{\reftag{Anton1})}{L. Antoniazzi et al., \PRL 70 (1993) 383.}
\par
\item{\reftag{Schuler})}{G. A. Schuler, ``Quarkonium Production and
Decays", CERN Preprint CERN-TH 7170/94, Feb. 1994.}
\par
\item{\reftag{KS5})}{D. Kharzeev and H. Satz, ``Charmonium Interaction
in Nuclear Matter", CERN Preprint CERN-TH/95--73, Apr. 1995.}
\par
\item{\reftag{Neubert})}{M. Beneke and V. M. Braun, ``Naive Non-Abelianisation
and Resummation of Fermion Bubble Chains", DESY Preprint DESY 94--200,
Nov. 1994. \hfill\break
M. Neubert, ``Scale Setting in QCD and the Momentum Flow in Feynman
Diagrams", CERN Preprint CERN-TH 7487/94, Dec. 1994}
\par
\item{\reftag{Badier})}{J. Badier et al., \ZP C 20 (1983) 101.)
\par
\item{\reftag{E772})}{D. M. Alde et al., \PRL 66 (1991) 133 and 2285.}
\par
\item{\reftag{Karsch})}{F. Karsch and H. Satz, \ZP C 51 (1991) 209.}
\par
\item{\reftag{Shadow})}{V. Heynen et al., \PL B 34 (1971) 651; \hfill \par
J. Bailey et al., \NP B151 (1979) 367; \hfill \par
for recent data, see \hfill\par
M. Arneodo et al., \PL B 211 (1988) 493;
\hfill \par
P. Amaudruz et al., \ZP C 51 (1991) 387.}
\par
\item{\reftag{Smirnov})}{G. I. Smirnov, ``A study of the A-dependence
of the deep inelastic scattering of leptons and its implications
for understanding of EMC effect", Dubna Preprint JINR-DI-94-278, 1994.}
\par
\item{\reftag{LP})}{L. D. Landau and I. Ya. Pomeranchuk,
Dokl. Akad. Nauk SSSR 92 (1953) 535 and 735;\hfill\par
E. L. Feinberg and I. Ya. Pomeranchuk,
Suppl.\ Nuovo Cim.\ III, Ser.\ X, No.\ 4 \hfill\par
(1956)
652;\hfill\par
M. L. Ter-Mikaelyan, J.E.T.P. 25 (1954) 289 and 296; \hfill \par
A. B. Migdal, \PR 103 (1956) 1811.}
\par
\item{\reftag{Lu})}{S.J. Brodsky and H. J. Lu, \PRL 64 (1990) 1342.}
\par
\item{\reftag{Gerschel})}{C. Gerschel and J. H\"ufner,
\ZP C 56 (1992) 171.}
\par
\item{\reftag{colour})}{S.J. Brodsky and A.H. Mueller,
\PL B 206 (1988) 685.}
\par
\item{\reftag{Farrar})}{G. Farrar et al., \PRL 64 (1990) 2996.}
\par
\item{\reftag{Blaizot})}{J.-P. Blaizot and J.-Y. Ollitrault,
\PL 217B (1989) 386.}
\par
\item{\reftag{EMC})}{J. J. Aubert et al., \NP B 293 (1987) 740.}
\par
\item{\reftag{Peskin})}{M. E. Peskin, \NP B156 (1979) 365.}
\par
\item{\reftag{Bhanot})}{G. Bhanot and M. E. Peskin, \NP B156 (1979)
391.}
\par
\item{\reftag{SVZ})}{M. A. Shifman, A. I. Vainshtein and V. I. Zakharov,
\PL 65B (1976) 255.}
\par
\item{\reftag{Novikov})}
{V. A. Novikov, M. A. Shifman, A. I. Vainshtein and V. I. Zakharov,
\NP B136 (1978) 125.}
\par
\item{\reftag{Kaidalov})}{A. Kaidalov, in {\sl QCD and High Energy
Hadronic Interactions}, J. Tran Thanh Van (Ed.), Editions Frontieres,
Gif-sur-Yvette, 1993.}
\par
\item{\reftag{Gross})}{D. J. Gross and F. Wilczek, \PR D8 (1973) 3633
and \PR D9 (1974) 980;\hfill\break
H. Georgi and H. D. Politzer, \PR D9 (1974) 416.}
\par
\item{\reftag{Parisi})}{G. Parisi, \PL 43B (1973) 207; 50B (1974) 367.}
\par
\item{\reftag{Povh})}{J. H\"ufner and B. Povh, \PRL 58 (1987) 1612.}
\par
\item{\reftag{MT})}{F. Karsch, M. T. Mehr and H. Satz, \ZP C 37 (1988) 617.}
\par
\item{\reftag{Larry})}{D. Kharzeev, L. McLerran and H. Satz,
``Non-Perturbative Quarkonium Dissociation in Hadronic Matter",
CERN Preprint CERN-TH/95-27, March 1995.}
\par
\item{\reftag{Low})}{F. E. Low, \PR D 12 (1975) 163.}
\par
\item{\reftag{GS2})}{S. Gupta and H. Satz, \PL B 283 (1992) 439.}
\par
\item{\reftag{Blaizot/Hwa})}{J.-P. Blaizot and J.-Y. Ollitrault in
{\sl Quark-Gluon Plasma}, R. C. Hwa (Ed.), World Scientific,
Singapore 1990).}
\par
\item{\reftag{counting})}{S. J. Brodsky and G. R. Farrar,
\PRL 31 (1973) 1153; \hfill\par
V. A. Matveev, R. M. Muradyan and A. N. Tavkhelidze, Nuovo Cimento
Lett. 7 (1973) 719.}
\par
\item{\reftag{splitting})}{G. Altarelli and G. Parisi, \NP B 126
(1977) 298.}
\par
\item{\reftag{Ronceux})}{B. Ronceux, Doctorate Thesis,
Universit\'e de Savoie, Annecy-le-Vieux, October 1993}
\par
\item{\reftag{Carlos})}{C. Lourenco, Doctorate Thesis,
Universidade T\'ecnica de Lisboa, Lisbon, January 1995}
\par
\item{\reftag{Gavin})}{S. Gavin et al., \ZP C 61 (1994) 351.}
\par
\item{\reftag{KS6})}{D. Kharzeev and H. Satz, ``\J~Suppression and
Pre-Equilibrium Colour Deconfinement",
CERN Preprint CERN-TH/95-120, May 1995.}
\par
\item{\reftag{Mueller})}{B. Mueller and K. Geiger, \NP B 369 (1992) 600.}
\par
\item{\reftag{Wang})}{M. Gyulassy and X.-N. Wang, \PR D44 (1992) 3501.}
\par
\item{\reftag{Shuryak})}{Li Xiong and E. V. Shuryak, \PR C 49 (1994)
2203.}
\par
\item{\reftag{NA38-new})}{C. Baglin et al., \PL B 345 (1995) 617.}
\vfill\eject\bye